\documentclass[journal]{IEEEtran}
\IEEEoverridecommandlockouts
\usepackage{cite}
\usepackage{amsmath}
\usepackage{graphicx}
\usepackage{multirow}
\usepackage{flushend}
\usepackage{acronym}
\usepackage{siunitx}
\usepackage{tabularx}
\usepackage{array}
\usepackage{ragged2e}
\usepackage[ruled,vlined]{algorithm2e}

\usepackage{algpseudocode}
\usepackage{booktabs}
\usepackage{longtable}
\usepackage{float}
\usepackage{balance}
\usepackage{subcaption}
\usepackage{tikz}
\usepackage{amsmath}
\usepackage{amssymb}
\usetikzlibrary{shapes.geometric, arrows.meta, positioning}

\acrodef{sdn}[SDN]{Software-Defined Networking}
\acrodef{ibn}[IBN]{Intent-Based Networking}
\acrodef{llm}[LLM]{Large Language Model}
\acrodef{llms}[LLMs]{Large Language Models}
\acrodef{mnrl}[MNRL]{Multiple Negatives Ranking Loss}
\acrodef{snn}[SNN]{Siamese Neural Network}
\acrodef{dt}[DT]{Digital Twin}
\acrodef{5g}[5G]{Fifth Generation}
\acrodef{6g}[6G]{Sixth Generation}
\acrodef{4g}[4G]{Fourth Generation}
\acrodef{ood}[OOD]{Out-of-Distribution}
\acrodef{id}[ID]{In-Distribution}
\acrodef{ran}[RAN]{Radio Access Network}
\acrodef{kpi}[KPI]{Key Performance Indicator}
\acrodef{nlp}[NLP]{Natural Language Processing}
\acrodef{urllc}[URLLC]{Ultra-Reliable Low-Latency Communication}
\acrodef{iot}[IoT]{Internet of Things}
\acrodef{nfv}[NFV]{Network Function Virtualization}
\acrodef{csc}[CSC]{Communication Service Customer}
\acrodef{csp}[CSP]{Communication Service Provider}
\acrodef{nop}[NOP]{Network Operator}
 \acrodef{3gpp}[3GPP]{Third Generation Partnership Project}

\def\BibTeX{{\rm B\kern-.05em{\sc i\kern-.025em b}\kern-.08em
    T\kern-.1667em\lower.7ex\hbox{E}\kern-.125emX}}

\usepackage[right=0.680in, left=0.680in, top=0.75in, bottom=1.02in]{geometry}

\begin{document}

\title{Semantic Routing Framework  for Multi-Domain Telecom Intent-Based Networking}

\author{
    {Marwa Tageldien,~\IEEEmembership{Student Member,~IEEE},
    Bassant Selim,~\IEEEmembership{Senior Member,~IEEE},
    and~Lokman Sboui,~\IEEEmembership{Senior Member,~IEEE}}
    \thanks{M.~Tageldien, B.~Selim, and L.~Sboui are with the Department
    of Systems Engineering, \'{E}cole de technologie sup\'{e}rieure, Montr\'{e}al,
    QC H3C 1K3, Canada (e-mail: marwa.tageldien@ieee.org;
    bassant.selim@etsmtl.ca; lokman.sboui@etsmtl.ca).}
}

\maketitle

% ============================================================
\begin{abstract}
Intent-Based Networking (IBN) is a key enabler of autonomous network
management in next-generation networks, as it allows operators to drive
the network through high-level intents instead of low-level configuration commands. In this paper, we propose a semantic routing framework for telecom intent classification and translation across the radio access networks (RANs), transport, 5G core, and cloud/edge domains. This framework operates locally to preserve intent privacy and avoid the latency and cost of cloud-based Large Language Model (LLM) solutions. The proposed framework is composed of intent classification and translation. To accurately classify the intent, we
propose fine-tuning sentence-transformer models using contrastive learning over a 3GPP
TS~28.312-aligned taxonomy of five top-level routes and 22 branches,
supported by a telecom-specific codebook of 1,100 intents. In addition, we develop a branch-aware, template-driven rule-based module for intent translation that enforces deterministic, hallucination-free parameter extraction and explicit missing-field detection, ensuring a structured and verifiable output pipeline. The numerical results show that the best fine-tuned encoder
achieves up to 99.02\%/94.61\% route/branch classification accuracy,
outperforming both base encoders and a commercial embedding baseline.
The proposed translator achieves up to 96.24\% translation accuracy,
compared to 85.37\% for a cloud-based LLM benchmark. Moreover, since the
proposed pipeline runs entirely on local hardware, an average translation
latency of 4.81~ms is recorded against 4,163.58~ms for the cloud-based
benchmark, i.e., a speedup of approximately 866$\times$.

\end{abstract}

% ============================================================
\section{Introduction}
\label{sec:intro}
\IEEEPARstart{I}{ntent}-based networking is a network management paradigm in which operators express \emph{what} the network should achieve through high-level, natural-language \emph{intents}, while the network itself determines \emph{how} to realize them~\cite{gharbaoui2023intent}. By decoupling management objectives from low-level configuration commands, \ac{ibn} reduces operational complexity and costs, accelerates service deployment, and enables autonomous network operation in beyond-\ac{5g} and \ac{6g} networks~\cite{wei2020intent, leivadeas2022survey, wang2025survey}. This vision has been embraced by both industry and standardization bodies. In particular, the \ac{3gpp} TS~28.312 specification defines intent-driven management services and formalizes how intents are expressed, interpreted, and fulfilled across the \ac{ran}, transport, 5G core, and edge domains~\cite{3gpp_ts28312_r18_v18_8_0}. As these domains grow in scale and heterogeneity, manual configuration becomes increasingly error-prone and costly, making intent-driven automation a practical necessity rather than a convenience.

\ac{ibn} research characterizes network management as a closed-loop process comprising intent profiling, translation, validation, activation, and assurance~\cite{gharbaoui2026assurance}. Among these lifecycle stages, intent classification in intent profiling identifies which management function an incoming intent belongs to, while intent translation converts it into a structured, machine-executable configuration. These two stages condition every downstream action, since a misclassified intent is dispatched to the wrong management function and an incompletely or wrong translated intent propagates into an incorrect network state. A practical intent classification and translation must therefore be accurate in the presence of natural-language variability, deterministic and verifiable in its output, and deployable under the privacy, latency, and cost constraints of real operator environments.
%%%%%%%%%%%%%%%%%%%%%%%%%%%%%%%%%%%%%
\subsection{Related Works}
\ac{ibn} has been studied across the full closed loop, e.g., through \ac{dt}-enabled interactive frameworks~\cite{tageldien2025gentwibn}, generative-AI-driven paradigms~\cite{qin2025generative}, and vision studies on intent-based slicing for \ac{iot} systems~\cite{hussein2025towards}. In this section, we focus on the prior work on intent classification and translation. Early studies addressing these two stages primarily relied on structured approaches, including controlled languages, predefined schemas, and rule-guided pipelines. For example, intents were expressed in a predefined format to drive slicing between containers over an \ac{sdn} overlay~\cite{rafiq2020intent}, exposed through a similar interface for P4-programmable data planes~\cite{riftadi2019p4i}, and used to steer multi-tenant Virtual Network Function (VNF) placement~\cite{leivadeas2021vnf}. Moving closer to natural language, the works in~\cite{jacobs2018refining} and \cite{jacobs2021hey} let operators type free-form sentences that are first mapped onto an intermediate intent representation and then refined into deployable policies. These approaches established the value of intermediate semantic representations and offered transparent, deterministic behavior, but they were mainly validated on small \ac{sdn} or enterprise testbeds and remained brittle under linguistic variability, requiring hand-written rules for each new phrasing or service. To relax this rigidity, subsequent methods augmented the intent text with additional meaning so that an intent can be recognized even when its wording departs from the predefined templates, using keyword extraction, knowledge graphs, and logical reasoning~\cite{xiao2022lightweight, alcock2025swift}. These approaches improve flexibility but still depend on manually crafted structures that are difficult to scale to operator-grade and multi-domain networks.

More recently, \acp{llm} have emerged as a promising solution for automating multiple stages of the \ac{ibn} lifecycle, owing to their strong natural-language understanding and reasoning capabilities~\cite{tageldien2025large}. The work in~\cite{manias2024towards} used an \ac{llm} to extract the key attributes of natural-language intents and classify them into management categories for the \ac{5g} core network. In~\cite{mekrache2024intent}, an \ac{llm}-centric architecture was proposed in which the \ac{llm} interprets operator intents and drives their fulfillment across next-generation network management. In addition, the work in~\cite{dzeparoska2023llm} employed an \ac{llm} to generate executable management policies directly from application-level intents, while~\cite{dinh2025towards} extended \ac{llm} assistance to the end-to-end intent lifecycle, from expression to deployment. In~\cite{asif2025evaluating}, several \acp{llm} were evaluated for intent translation, complemented by a k-nearest-neighbor mechanism to detect contradictions between newly issued and existing intents. The work in~\cite{manias2024semantic} introduced a semantic routing layer that classifies an intent into routes by embedding similarity before forwarding it to the \ac{llm}, improving both efficiency and accuracy. These works demonstrate that \acp{llm} handle paraphrases and unseen formulations far better than rule-based systems, and a parallel line of work has emphasized that trust and verifiability are prerequisites for natural-language-driven management~\cite{gharbaoui2026assurance, jacobs2025establishing}. Nevertheless, as detailed in the following subsection, each family of approaches, whether rule-based, semantically enriched, or \ac{llm}-based, exhibits limitations that are incompatible with the requirements of practical \ac{ibn} deployments, and these limitations motivate the present work.
%%%%%%%%%%%%%%%%%%%%%%%%%%%%%%%%%%%%%%%%%%%%%%%%%
\subsection{Motivation}
The intent classification and translation stages condition every downstream action of the \ac{ibn} loop, so their limitations propagate directly into network behavior. Examining the related works from this perspective reveals the following limitations. Rule-based and semantically enriched approaches rest on controlled languages, fixed schemas, or hand-crafted ontologies, so they reject or misinterpret any intent whose wording departs from the anticipated form. Each new phrasing, service, or domain requires additional expert-written rules, which limits their usability and scalability in operator-grade networks. \ac{llm}-based approaches overcome this variability but depend on remote, third-party \ac{llm} APIs. Hence, every intent is transmitted to an external provider, which is unacceptable for private, enterprise, and mission-critical networks whose intents disclose sensitive topology and policy information. Additionally, the multi-second round-trip violates the millisecond-level budgets of closed-loop use cases such as self-healing activation and \ac{urllc} management. Furthermore, since these approaches translate intents through open-ended text generation, a hallucinated parameter, an omitted mandatory field, or a non-canonical attribute name results in an invalid configuration~\cite{asif2025evaluating, wei2025inta}. Additionally, the output varies across repeated executions of the same intent, which prevents the machine-checkable validation required before deployment~\cite{gharbaoui2026assurance, jacobs2025establishing}.

Beyond these method-specific issues, existing taxonomies and datasets are built around \ac{sdn}, enterprise, or \ac{5g} core use cases and classify intents at a single, coarse level. \ac{ran} management, however, is governed by radio resource management, coverage, interference, and mobility requirements with no direct equivalent in the core. So, a core-centric classifier misroutes or rejects \ac{ran}-specific intents, while a single-level classifier cannot separate intents that share a management objective but differ in the requested operation. Finally, the absence of publicly available, \ac{3gpp}-aligned intent datasets prevents fair benchmarking and forces reliance on general-purpose encoders that lack awareness of telecom terminology and \ac{3gpp} management concepts.

These limitations reveal four research gaps in practical \ac{ibn} deployment: robustness to natural-language variability, local and low-latency execution, deterministic and verifiable translation, and multi-domain, fine-grained coverage aligned with the \ac{3gpp} standards. 
%To meet them, we propose a semantic routing framework based on the \ac{3gpp} TS~28.312-aligned route/branch taxonomy, locally fine-tuned sentence-transformer models for intent classification, and a deterministic template-driven translation module, together with a reproducible, 3GPP-aligned intent dataset.

%%%%%%%%%%%%%%%%%%%%%%%%%%%%%
\subsection{Contributions}
\begin{figure}[t]
\includegraphics[width=\linewidth]{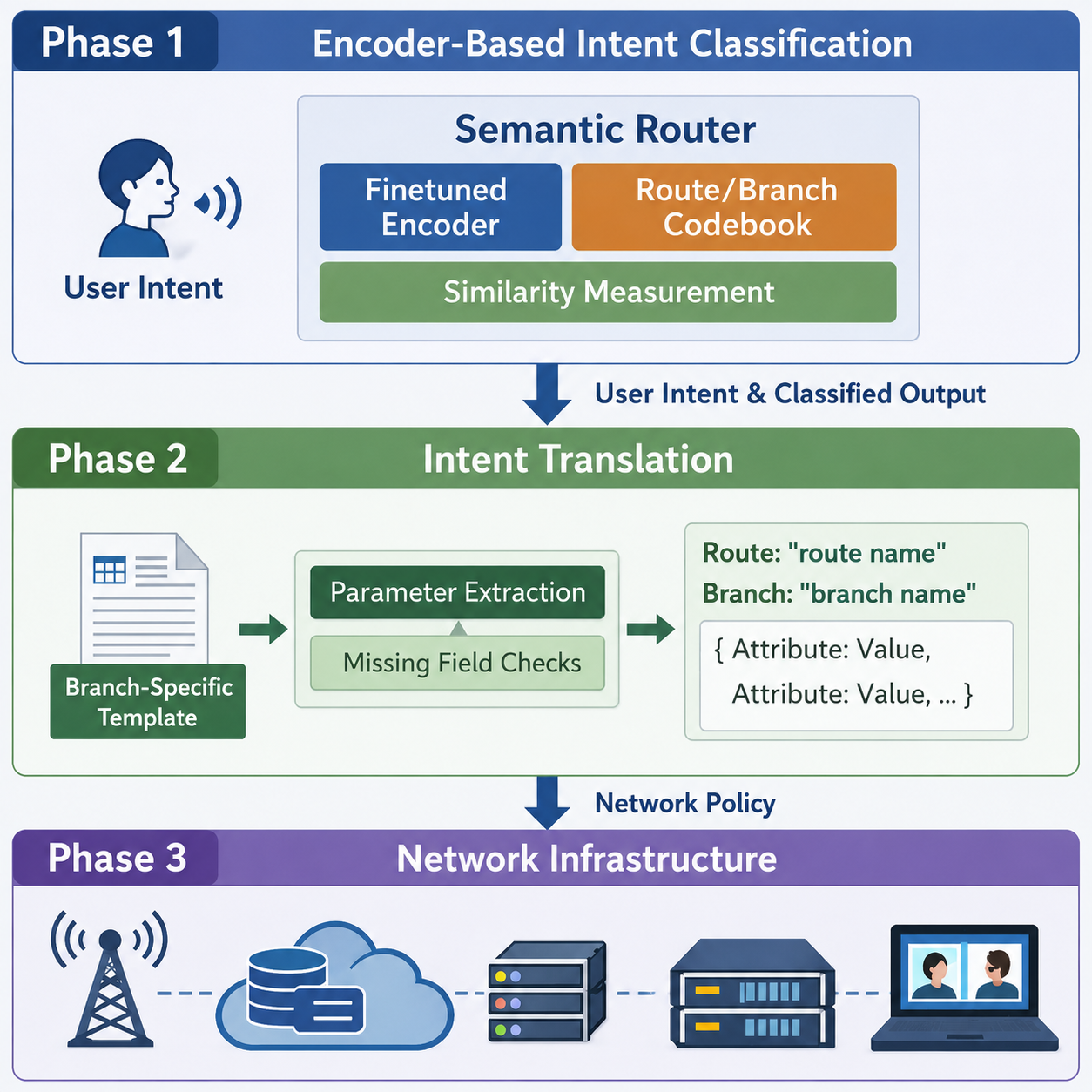}
\caption{Proposed semantic routing framework for multi-domain telecom intent-based networking.}
\label{fig:system_model}
\end{figure}

To address the identified research gaps, we propose a semantic routing framework for multi-domain telecom intent classification and translation aligned with the \ac{3gpp} TS~28.312 (V18.8.0) specification~\cite{3gpp_ts28312_r18_v18_8_0}, designed for local and low-latency deployment. As shown in Fig.~\ref{fig:system_model}, the proposed framework first employs locally fine-tuned sentence-transformer encoders to map natural-language intents into a structured route/branch taxonomy, followed by a branch-aware, template-driven translation module. The main contributions of this work are summarized as follows:

\begin{itemize}
\item We propose a semantic routing framework for telecom intent classification and translation that runs entirely on local hardware. By replacing remote \ac{llm} APIs with locally fine-tuned sentence-transformer encoders, the framework limits inference latency and computational cost while preserving intent privacy.

\item We design a two-level route/branch taxonomy inspired by \ac{3gpp} TS~28.312 (V18.8.0), consisting of five top-level routes and 22 branches, which classifies the management function and the specific operation hierarchically and covers the \ac{ran}, transport, \ac{5g} core, and edge domains.

\item We construct a telecom-specific codebook of 1,100 intents enriched using five linguistic variation methods. Derive from it a reproducible dataset for contrastive fine-tuning and evaluation of sentence-transformer encoders, providing robustness to natural-language variability and enabling standardized benchmarking for telecom intent classification.

\item We develop a branch-aware, template-driven intent translation mechanism based on deterministic parameter extraction, ensuring schema compliance, explicit missing-field detection, and the elimination of hallucinated outputs.
\end{itemize}

The remainder of this paper is organized as follows: Section~\ref{sec:system} presents the system model. Sections~\ref{sec:classification} and~\ref{sec:translation} describe the encoder-based intent classification and translation modules, respectively. Section~\ref{sec:experiments} presents the results, and Section~\ref{sec:conclusion} concludes this work.

% ============================================================
\section{System Model}
\label{sec:system}

% Justified tabularx column (no paragraph indent)
\newcolumntype{J}{>{\justifying\setlength{\parindent}{0pt}\arraybackslash}X}

\begin{table}[t]
\centering
\caption{User types with intent examples from the 3GPP TS28.312 specifications \cite{3gpp_ts28312_r18_v18_8_0}.}
\label{tab:user_types}
\renewcommand{\arraystretch}{1.15}
\setlength{\tabcolsep}{3.8pt}
\small

% Col1 fixed, Col2 fixed (reduced), Col3 flexible (increased)
\begin{tabularx}{\columnwidth}{|p{0.22\columnwidth}|p{0.33\columnwidth}|J|}
\hline
\textbf{User Type} & \textbf{Definition} & \textbf{3GPP Intent Examples} \\ \hline

Communication Service Customer (CSC) &
Expresses an intent that specifies the communication service outcome to be consumed, without specifying the underlying service management or network realization details. &
\textit{"Enable a V2X communication service for a group of vehicles during a specified time window."} \\ \hline

Communication Service Provider (CSP) &
Expresses an intent that specifies the network service outcome to be delivered, without specifying the detailed network management actions required for implementation. &
\textit{"Provide a network service supporting V2X communications on Highway-417 for up to 500 vehicles simultaneously."} \\ \hline

Network Operator (NOP) &
Expresses an intent that specifies the subnetwork management objective for a group of network elements, without specifying the detailed element-level procedures or configuration steps. &
\textit{"Provide a radio network service that meets the specified coverage and UE throughput requirements in a target area."} \\ \hline

\end{tabularx}
\end{table}

This section presents the proposed semantic routing framework for telecom encoder-based intent classification and translation. As shown in Fig.~\ref{fig:system_model}, the framework classifies a user-defined natural-language intent through the encoder-based intent classification stage, producing a classified output consisting of the corresponding route and branch. The classified intent is then transformed by the intent translation stage into a structured, validated, and machine-executable network policy. The resulting policy is subsequently deployed directly within the network infrastructure. The network infrastructure comprises the underlying network devices and resources, including routers, switches, base stations, etc.

The proposed framework follows the intent-driven management architecture specified in the \ac{3gpp} TS~28.312 specifications ~\cite{3gpp_ts28312_r18_v18_8_0}, which defines how intents are expressed, interpreted, and translated into network management actions. According to this specification, \ac{ibn} systems support heterogeneous users with different responsibilities and levels of technical expertise. As summarized in Table~\ref{tab:user_types}, three primary user types are defined: the \ac{csc}, the \ac{csp}, and the \ac{nop}. These users express management objectives using natural-language intents at different abstraction levels, ranging from service-oriented requests for the \ac{csc} to detailed operational directives in \ac{nop} scenarios. In this work, without loss of generality, we consider the \ac{nop} as the primary user. Accordingly, the framework accepts
high-level \ac{nop} intents as input and automatically transforms them
into structured network policies. Nevertheless, the proposed framework
can be extended to \ac{csc} and \ac{csp} intents by enriching
the codebook and the associated branch templates with utterances at the
corresponding abstraction levels.

In this work, we consider a multi-domain telecom network encompassing
the \ac{ran}, the transport network, the 5G core network, and edge
computing infrastructure, consistent with the scope of the intent-driven
management services defined in 3GPP TS~28.312. Each domain exposes a collection of managed entities, including cells, network slices, transport links, core network functions, and edge service instances, each associated with a specific geographical or operational scope. These entities are continuously monitored through standardized \acp{kpi}, such as throughput, latency, coverage, reliability, and resource utilization. Consequently, every intent targeting  network management request can be represented using a common set of attributes comprising the target entity, operational scope, requested management action, performance objectives, operational constraints, and temporal context. These common attributes provide a unified representation of network management intents across heterogeneous telecom domains. In addition to sharing a common attribute structure, the intents issued over these managed entities correspond to a recurring set of management tasks, including the configuration lifecycle of network services, the feasibility and conflict resolution of requested configurations,
performance assurance against service-level targets, autonomous
optimization and healing loops, and analytics and operational reporting.
Notably, each of these network management operations can be applied to any network segment. For example, a performance monitoring request may
target a \ac{ran} cell, a transport link, or a 5G core network function.
In other words, the type of management operation and the network segment
on which it acts constitute two independent dimensions of an intent.
This observation motivates the two-level route/branch taxonomy presented
in Section~\ref{sec:classification}.

%In what follows, we present the main components of the proposed semantic routing framework, including an encoder-based intent classification, intent translation, and the numerical results of the system within the \ac{ibn} workflow.
%%%%%%%%%%%%%%%%%%%%%%%%%%
\section{Phase 1: Proposed Encoder-based Intent Classification}
\label{sec:classification}
This section presents the proposed encoder-based intent classification,
in which fine-tuned sentence-transformer encoders map an input intent
to its route/branch label through semantic similarity matching.
The design of this classifier follows three steps. First, we define a
3GPP-aligned route/branch taxonomy for intent representation.
Second, we build a telecom codebook and dataset using \ac{llm}-based
generation and linguistic diversity methods. Third, we fine-tune
sentence-transformer encoders with contrastive learning for accurate
intent classification. The remainder of this section details each of
these steps.
%%%%%%%%%%%%%%%%%%%%%%%%%%%%%%%%%%%
\subsection{Routes/Branches Taxonomy}
The proposed encoder-based intent classification (phase 1), illustrated in Fig. \ref{fig:system_model}, organizes the routes/branches taxonomy as a two-level classification: routes and branches.
The route represents a top-level network management category that determines
which management function should process the incoming intent. While the branch refines
the selected route into a specific operation within that management category. The idea of proposing two-classification levels, route/branch structure, is motivated by the need to reduce the complexity of intent understanding in large-scale telecom systems. Direct classification over a large number of intent classes is
challenging due to the high semantic overlap between classes and the
variability of natural-language formulations. By first mapping intents
into a small set of high-level management routes, the search
space is significantly constrained, improving both classification
robustness and scalability.
% One-column table, wider Route Name column (take space from Definition)
\begin{table}[t]
\centering
\caption{Route Names.}
\label{tab:routes}
\renewcommand{\arraystretch}{1.15}
\setlength{\tabcolsep}{4pt}
\small
\begin{tabular}{|c|p{0.30\columnwidth}|p{0.48\columnwidth}|}
\hline
\textbf{Route \#} & \textbf{Route Names (R)} & \textbf{Definition} \\ \hline
1 & Network Configuration Management
  & Focuses on the management of network configurations from creation to deletion. \\ \hline
2 & Configuration Feasibility Management
  & Ensures that requested configurations are feasible, conflict-free, and aligned with resources before implementation. \\ \hline
3 & Performance Assurance \& Compliance
  & Monitors the network's operational quality against defined standards and metrics. \\ \hline
4 & Autonomous Network Optimization \& Healing
  & Manages the self-governing capabilities of the network, allowing it to adapt and fix issues without human intervention. \\ \hline
5 & Network Analytics \& Operational Reporting
  & Provides the necessary functions to retrieve data, generate reports, and gain insights into the network's current state and history. \\ \hline
\end{tabular}
\end{table}

Guided by this design principle, we propose a two-level taxonomy
inspired by the 3GPP TS~28.312 (V18.8.0)
specifications, which define the
intent-driven management services for mobile networks. Additionally, these specifications describe
representative use cases spanning the configuration, feasibility
checking, performance assurance, autonomous optimization, and reporting
of network services. We group these recurring use cases into five
top-level routes, each corresponding to a standardized telecom
management function, while the intent examples and attributes provided
in the specification guide the construction of the branch templates and
codebook utterances presented in the subsequent subsections.
Accordingly, we define five top-level routes: \textit{Network Configuration Management (R1)}, \textit{Configuration Feasibility Management (R2)}, \textit{Performance Assurance \& Compliance (R3)}, \textit{Autonomous Network Optimization \& Healing (R4)}, and \textit{Network Analytics \& Operational Reporting (R5)}, as summarized in Table~\ref{tab:routes}. Each route is further decomposed into branches that capture recurring operational patterns within that domain. These branches are summarized in Table~\ref{tab:all_routes_branch_templates} with a total of 22 branches. This hierarchical decomposition enables fine-grained discrimination between semantically similar intents that differ in operational detail rather than high-level objective. Each branch is associated with a base utterance template that specifies the minimal structural elements required to identify the management intent, including the action verb, entity identifier, operational scope, and, where applicable, performance targets and constraints. To illustrate the mapping, consider the following example intent: \textit{``Update
network configuration NR-MTL-DTN-n258 in Montreal Downtown by adjusting throughput
targets and latency constraints to achieve average DL RAN UE throughput $\geq$
140\,Mbps and DL latency $\leq$ 12\,ms.''}. In this case, the action verb \textit{"update"} places the intent under (\textit{R1: Network Configuration Management}), while the structure of the request—modifying an existing configuration to meet updated performance objectives—matches (\textit{B3: Update Network Configuration Parameters}). Hence, the intent is classified as R1,B3.

\begin{table*}[t]
\centering
\caption{Route Branches and Their Base Utterance Templates.}
\label{tab:all_routes_branch_templates}

\renewcommand{\arraystretch}{1.25}
\setlength{\tabcolsep}{7.5pt}
\small

\begin{tabular}{|c|c|p{0.23\textwidth}|p{0.58\textwidth}|}
\hline
\textbf{R \#} & \textbf{B \#} & \textbf{Branch Name (B)} & \textbf{Base Utterance} \\
\hline

\multirow{5}{*}{\textbf{1}}
& \textbf{B1} & Define Network Configuration &
``Define a new network configuration [entity identifier] in [scope] with [parameters]". \\ \cline{2-4}

& \textbf{B2} & Activate Network Configuration &
``Activate [entity identifier] in [scope] during [temporal context]''. \\ \cline{2-4}

& \textbf{B3} & Update Network Configuration Parameters &
``Update [entity identifier] in [scope] by changing [parameters]". \\ \cline{2-4}

& \textbf{B4} & Stop Configuration Execution &
``Stop execution of [entity identifier] in [scope] during [temporal context]". \\ \cline{2-4}

& \textbf{B5} & Delete Network Configuration &
``Delete [entity identifier] from [scope] under [constraints]". \\ \hline

\multirow{4}{*}{\textbf{2}}
& \textbf{B1} & Assess Network Feasibility &
``Assess feasibility of [entity identifier] in [scope] to meet [targets] under [constraints]". \\ \cline{2-4}

& \textbf{B2} & Set Priority &
``Set priority of [entity identifier] to [priority level] in [scope] during [temporal context]". \\ \cline{2-4}

& \textbf{B3} & Check Conflict Status &
``Check conflicts between [entity identifier] and [entity identifier2] in [scope], and report conflicts". \\ \cline{2-4}

& \textbf{B4} & Resolve Conflict &
``Resolve conflict between [entity identifier] and [entity identifier2] in [scope] by adjusting [parameters/scope/schedule] under [constraints]". \\ \hline

\multirow{4}{*}{\textbf{3}}
& \textbf{B1} & Performance Monitoring &
``Monitor [KPIs] for [entity identifier] in [scope] every [frequency] during [temporal context]". \\ \cline{2-4}

& \textbf{B2} & Evaluate KPI/SLA Compliance &
``Evaluate whether [entity identifier] in [scope] meets [targets] over [temporal context], and report compliance". \\ \cline{2-4}

& \textbf{B3} & Trigger Performance Deviation Alerts &
``Trigger if [KPI] violates [threshold] for [duration] for [entity identifier] in [scope]". \\ \cline{2-4}

& \textbf{B4} & Forecast Performance Compliance &
``Forecast whether [entity identifier] in [scope] remains compliant with [targets] over the next [duration]". \\ \hline

\multirow{4}{*}{\textbf{4}}
& \textbf{B1} & Configure Self-Optimization Control Loop &
``Enable optimization loop [entity identifier] in [scope] every [frequency] to achieve [targets] under [constraints]". \\ \cline{2-4}

& \textbf{B2} & Configure Self-Healing Actions &
``Enable self-healing for [entity identifier] in [scope] when [condition] holds for [duration], apply [actions] to restore [targets]". \\ \cline{2-4}

& \textbf{B3} & Set Adaptation Triggers &
``Trigger adaptation for [entity identifier] in [scope] when [predicted/observed condition] holds for [duration]". \\ \cline{2-4}

& \textbf{B4} & Disable Autonomous Control Loop &
``Disable the control loop [entity identifier] in [scope] during [temporal context]". \\ \hline

\multirow{5}{*}{\textbf{5}}
& \textbf{B1} & Query Operational State &
``Query the current operational state of [entity identifier] in [scope], including [state fields]". \\ \cline{2-4}

& \textbf{B2} & Retrieve Metric Data &
``Retrieve [metrics] for [scope] over [temporal context] with [granularity]". \\ \cline{2-4}

& \textbf{B3} & Retrieve Fulfillment Actions Log &
``Retrieve the actions log for [entity identifier] in [scope] over [temporal context], including [fields]". \\ \cline{2-4}

& \textbf{B4} & Generate Network Performance Report &
``Generate a performance report for [entity identifier] in [scope] over [temporal context], summarizing [targets]". \\ \cline{2-4}

& \textbf{B5} & List Managed Network Services/Configurations &
``List managed [services/configurations] in [scope], optionally filtered by [criteria]". \\ \hline

\end{tabular}
\end{table*}

%%%%%%%%%%%%%%%%%%%%%%%%%%%%%%%%
\subsection{Codebook and Dataset Generation}
\label{sec:dataset}
%\paragraph{Codebook Construction}
Fine-tuning the sentence-transformer encoders for intent classification
requires a labeled set of intent samples spanning all
route/branch categories. Since no publicly available intent
dataset aligned with the 3GPP specifications exists, we design a
telecom-specific codebook and construct the training dataset from its
content. The codebook is a structured repository of natural-language
intent samples, known as \textit{utterances}, where each utterance
is mapped to a unique route/branch label. It is organized according to the route/branch taxonomy: each route
contains its branches, and each branch is populated
with a set of utterances. The codebook utterances are created through a
two-stage process: base intent generation and linguistic diversity
methods.
%%%%%%%%%%%%%%%%%%%%%%%%
\paragraph{Base Intent Generation}
For each branch, ten base intents are generated using \acp{llm} through prompt engineering, guided by the eight intent-driven management use cases described in 3GPP TS~28.312 (V18.8.0). The prompts are designed with two objectives. First, each base intent must contain the minimum structural elements needed for route/branch classification, following the base utterance template of its branch. Second, the generated intents cover
the different network segments: some intents target a single segment, such as the \ac{ran}, the core network, or an edge service, while others describe end-to-end scenarios involving multiple segments simultaneously, such as an end-to-end network slice. All generated intents are manually reviewed and refined to ensure
alignment with the 3GPP specifications and to remove ambiguity in route/branch mapping.
%%%%%%%%%%%%%%%%%%%%%%%
\paragraph{Linguistic Diversity Methods}
To ensure that the fine-tuned sentence-transformer encoders are robust to natural phrasing variability,
five linguistic diversity methods are applied to each base intent. Specifically, we employ two prompting strategies
inspired from \cite{manias2024semantic}, namely \textit{Variability Prompt} and \textit{Paraphrased Prompt}, and add the following three proposed methods:

\begin{itemize}
    \item \emph{Variability Prompting}: Introduces lexical diversity by adjusting
    wording and phrasing while preserving the original sentence structure and semantic
    intent \cite{manias2024semantic}.
    \item \emph{Paraphrased Prompting}: Introduces stronger diversity by modifying both
    wording and sentence structure while maintaining semantic equivalence \cite{manias2024semantic}.
    \item \emph{Formality/Register Variation}: Generates three versions of each base
    intent in formal, neutral, and informal registers, adapting vocabulary and tone
    while preserving meaning. We use the
    following prompt to generate the variant intents: “Take the following base intent and rewrite it in three different registers: (a) formal, (b) neutral, and (c) informal. Make sure the meaning stays consistent, but adapt the vocabulary, tone, and sentence structure accordingly.”
    \item \emph{Active/Passive Voice Transformation}: Produces active-voice and passive-voice realizations of each base intent without altering the requested
    action, entities, or urgency. We utilize the following prompt for this method: "Take the following base intent and rewrite it in the active and passive voice. Keep the intent meaning and label each version clearly.”
    \item \emph{Negation/Contrast Reformulation}: Rephrases each base intent using
    negated or contrastive formulations (e.g., prohibitions, consequence-based framing)
    that imply the same required management action. We use the following prompt for this method: “Take the following base intent and rephrase it using a negative or
contrastive formulation that conveys the same intent. Try two different ways: (a) direct negation, and (b) implied consequence or contrast.”
\end{itemize}

Applying these five methods to each of the ten base intents per branch yields 50 utterances per branch and a total of 1,100 base and diverse intents across the 22 branches.
%%%%%%%%%%%%%%%%%%%%%%%%%%%%%%%%%%
\paragraph{Training Dataset Construction}
To fine-tune the sentence transformer encoders, we create two datasets, constructed from the codebook utterances, to support
fine-tuning with two different contrastive learning loss functions. Contrastive learning is a family of self-supervised and supervised representation
learning methods that shapes an embedding space by drawing similar samples together and
pushing dissimilar samples apart~\cite{chen2020simple, hu2024comprehensive}. Given a
set of training samples, a contrastive loss function penalizes the encoder when
semantically equivalent samples produce dissimilar embeddings, or when non-equivalent
samples produce similar embeddings.

In this work, we utilize two contrastive loss functions for the two datasets: \ac{mnrl} and Triplet loss.  The first constructed dataset (\texttt{Anchor-Positive-Pairs.jsonl}) contains anchor--positive
pairs, where both the anchor and the positive are utterances from the same branch. We utilize the \ac{mnrl} with this dataset because it operates on anchor--positive pairs $(A_i, P_i)$, where the anchor and
positive are semantically equivalent utterances from the same branch. Within a
mini-batch of size $B$, the positive $P_i$ is treated as the correct match for anchor
$A_i$, while all other positives $\{P_j\}_{j \neq i}$ are used as implicit in-batch
negatives. As  shown in 
equation~(\ref{eq:cross}), the \ac{mnrl} loss function maximizes the similarity of the correct pair relative to
the full batch using a cross-entropy formulation ~\cite{taneja2023efficient}, where $V_{A_i} = f_\Phi(A_i)$ and $V_{P_j} = f_\Phi(P_j)$ are the embedding vectors
of the anchor and the $j$-th positive, respectively. \ac{mnrl} implicitly exposes the
encoder to a large number of competing negatives without requiring explicit negative
sampling, making it computationally efficient and well-suited to the semantic router
retrieval setting.  So, under the \ac{mnrl}, all other positives in the same mini-batch serve
as implicit negatives.
\begin{equation}
L_{\text{MNRL}} = -\frac{1}{B} \sum_{i=1}^{B}
  \log \frac{\exp S(V_{A_i}, V_{P_i})}{\sum_{j=1}^{B} \exp S(V_{A_i}, V_{P_j})}
\label{eq:cross}
\end{equation}

The second dataset (\texttt{Anchor-Positive\-Negative-Pairs.jsonl}) contains triplets,
where the anchor and positive are from the same branch, and the negative is sampled from
a different branch within the same route. Hence, for this dataset we utilize the triplet loss function since it operates on triplets $(A, P, N)$, where the anchor $A$ and positive $P$
belong to the same branch, and the negative $N$ is drawn from a different branch within
the same route. As shown in equation (\ref{eq:triple}), the triplet loss function enforces a margin $\alpha$ between the anchor--positive
similarity and the anchor--negative similarity. By selecting negatives from the same route as the anchor, the triplets are constructed to be semantically hard, since different branches within the same route share coarse-grained semantics while differing in the specific requested operation. This
design aligns triplet supervision with the route/branch classification
objective.
\begin{equation}
L_{\text{triplet}} = \max\!\left(0,\;
  S(V_A, V_N) - S(V_A, V_P) + \alpha\right)
  \label{eq:triple}
\end{equation}
%%%%%%%%%finetuning figure%%%%%%%%%%%%
\begin{figure*}[h]
\includegraphics[width=\linewidth]{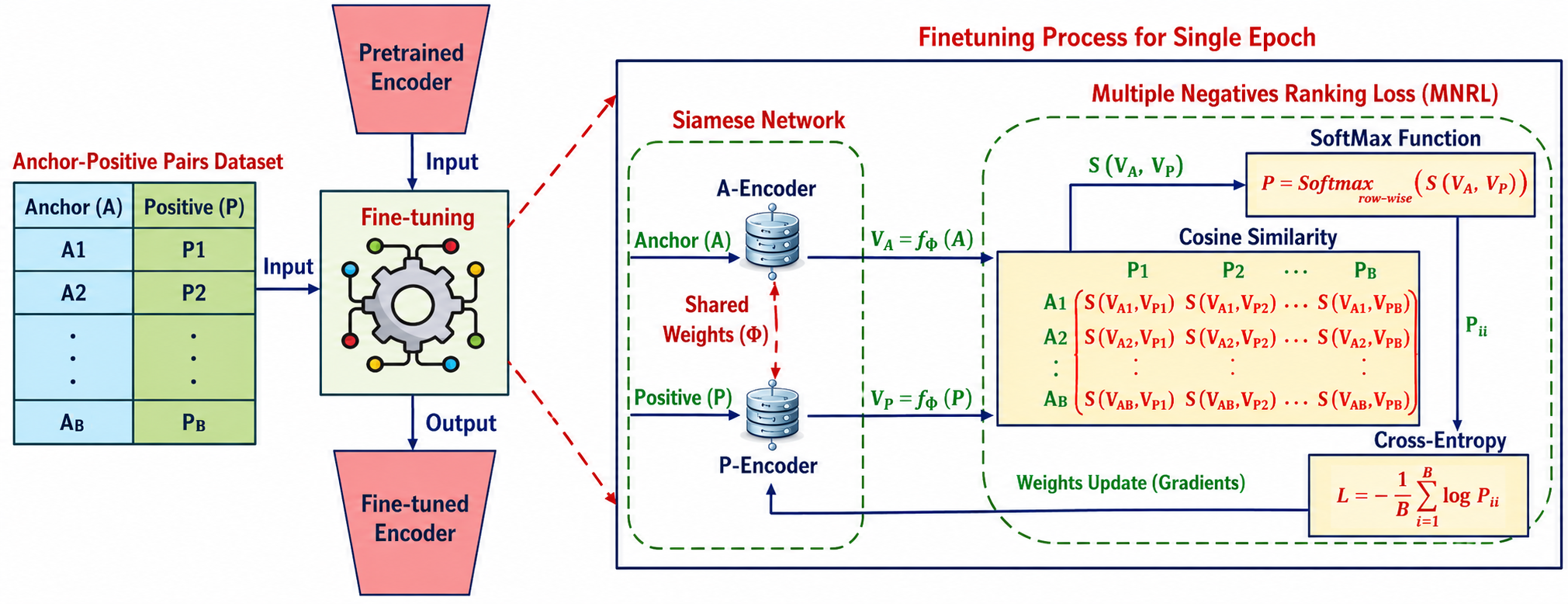}
\caption{Fine-tuning process for the sentence-transformer models for a single epoch.}
\label{fig:tuning}
\end{figure*}
In this work, intra-route negatives are preferred over
cross-route negatives because they are semantically harder. Branches within the same
route share coarse-grained semantics while differing in the specific requested operation, providing more informative training signals. The total number of training samples across all branches is calculated using equation (\ref{eq:pair_count}), where $N_i$ is the number of utterances in branch $i$.
\begin{equation}
\text{Total-Training-Pairs} = \sum_{i \in \text{branches}} N_i \left(N_i - 1\right)
\label{eq:pair_count}
\end{equation}
%%%%%%%%%%%%%%%%%%%%%%%%%%%%%%%%%%%%%
%The first constructed dataset (\texttt{Anchor-Positive-Pairs.jsonl}) contains anchor--positive pairs, where both the anchor and the positive are utterances from the same branch. Under \ac{mnrl}, all other positives in the same mini-batch serve as implicit negatives. The second dataset (\texttt{Anchor-Positive\-Negative-Pairs.jsonl}) contains triplets, where the anchor and positive are from the same branch, and the negative is sampled from a different branch within the same route. Intra-route negatives are preferred over cross-route negatives because they are semantically harder: branches within the same route share coarse-grained semantics while differing in the specific requested operation, providing more informative training signals. The total number of the training samples across all \textit{branches} is calculated using equation (\ref{eq:pair_count}), where $N_i$ is the number of utterances in branch $i$.
%\begin{equation}
%\text{Total-Training-Pairs} = \sum_{i \in \text{branches}} N_i \left(N_i - 1\right)
%\label{eq:pair_count}
%\end{equation}

%%%%%%%%%%%%%%%%%%%%%%%%%%%%%%%%%
\subsection{Fine-Tuning of Sentence-Transformer Models}
\label{sec:finetuning}

This subsection presents the fine-tuning of pretrained sentence-transformer models to the \ac{ibn}
domain. In this work, the fine-tuning process depends on the sentence-transformer models, \ac{snn}, and the contrastive loss function. The sentence-transformer models are pretrained transformer-based models adapted for
producing fixed-dimensional sentence embeddings suitable for semantic similarity
tasks~\cite{li2022survey}. Unlike standard language models that produce contextual
token embeddings, sentence transformers apply a pooling strategy (e.g., mean pooling)
over the token representations to yield a single embedding vector per input
sentence~\cite{colangelo2025comparative}. Without loss of generality, two sentence-transformer models are employed in this work: \texttt{all-mpnet-base-v2}~\cite{hussain2025transforming}, which produces 768-dimensional embeddings and is optimized for high
semantic accuracy, and \texttt{all-MiniLM-L6-v2}~\cite{yin2024study}, which produces 384-dimensional embeddings and prioritizes inference
efficiency. These two models are selected as they are among the best-performing open-source encoders on public embedding benchmarks ~\cite{song2020mpnet}. In addition, they represent the two ends of the accuracy--efficiency trade-off, matching the accuracy-oriented and latency-constrained deployment scenarios considered in this work. These two models are pretrained on general-domain text corpora, and therefore lack awareness of telecom-specific terminology, 3GPP management concepts, and the structural conventions of network management intents. This domain gap
motivates the fine-tuning procedure, which adapts the embedding space to the \ac{ibn}-specific route/branch taxonomy through contrastive training.

Besides the two sentence-transformer models, the fine-tuning process is based on a  \ac{snn}, which is well suited for representation learning from semantically related text pairs. In the \ac{snn}, two input utterances are processed in parallel by identical sub-networks that share the same parameters, and the resulting
embeddings are compared using a similarity measure \cite{li2022survey}. This structure enables the encoder to learn an embedding space in which semantically equivalent intents are placed closer together, while non-matching intents are pushed farther apart. In this work, the sentence-transformer encoders are fine-tuned to learn how to generate the embedding vector for
the input intent.

For fine-tuning we utilize the \ac{mnrl} and triplet loss functions but in what follows, we detail the fine-tuning process using the \ac{mnrl} loss function. As illustrated in Fig.~\ref{fig:tuning}, the fine-tuning input consists of anchor--positive pairs drawn from the training dataset, together with a pretrained
sentence-transformer encoder. The Anchor ($A$) is a base intent utterance, and the
Positive ($P$) is a linguistic diversity intent from the same branch. The \ac{snn} comprises
two encoders, the A-Encoder and the P-Encoder, sharing parameters $\Phi$. As shown in equation \ref{eq:v}, the
embedding vectors are computed as:
\begin{equation}
\label{eq:v}
V_A = f_{\Phi}(A), \qquad V_P = f_{\Phi}(P).
\end{equation}
For a mini-batch of size $B$, pairwise cosine similarities are computed between all
anchor and positive embeddings, as shown in equation (\ref{eq:sim}). Then, a row-wise Softmax converts each row of the similarity matrix into a probability
distribution, as  illustrated in equation (\ref{eq:soft}), where $P_{ii}$ is the probability assigned to the correct 
$(A_i, P_i)$ pair. The \ac{mnrl} loss (\textit{L}) is then calculated as shown in equation (\ref{eq:mnrl}).
\begin{equation}
S(V_{A_i}, V_{P_j}) = \cos(V_{A_i}, V_{P_j}), \quad S \in \mathbb{R}^{B \times B}.
\label{eq:sim}
\end{equation}
%%%%%%%%%Semantic Router figure%%%%%%%%%
\begin{figure*}[t]
\includegraphics[width=\linewidth]{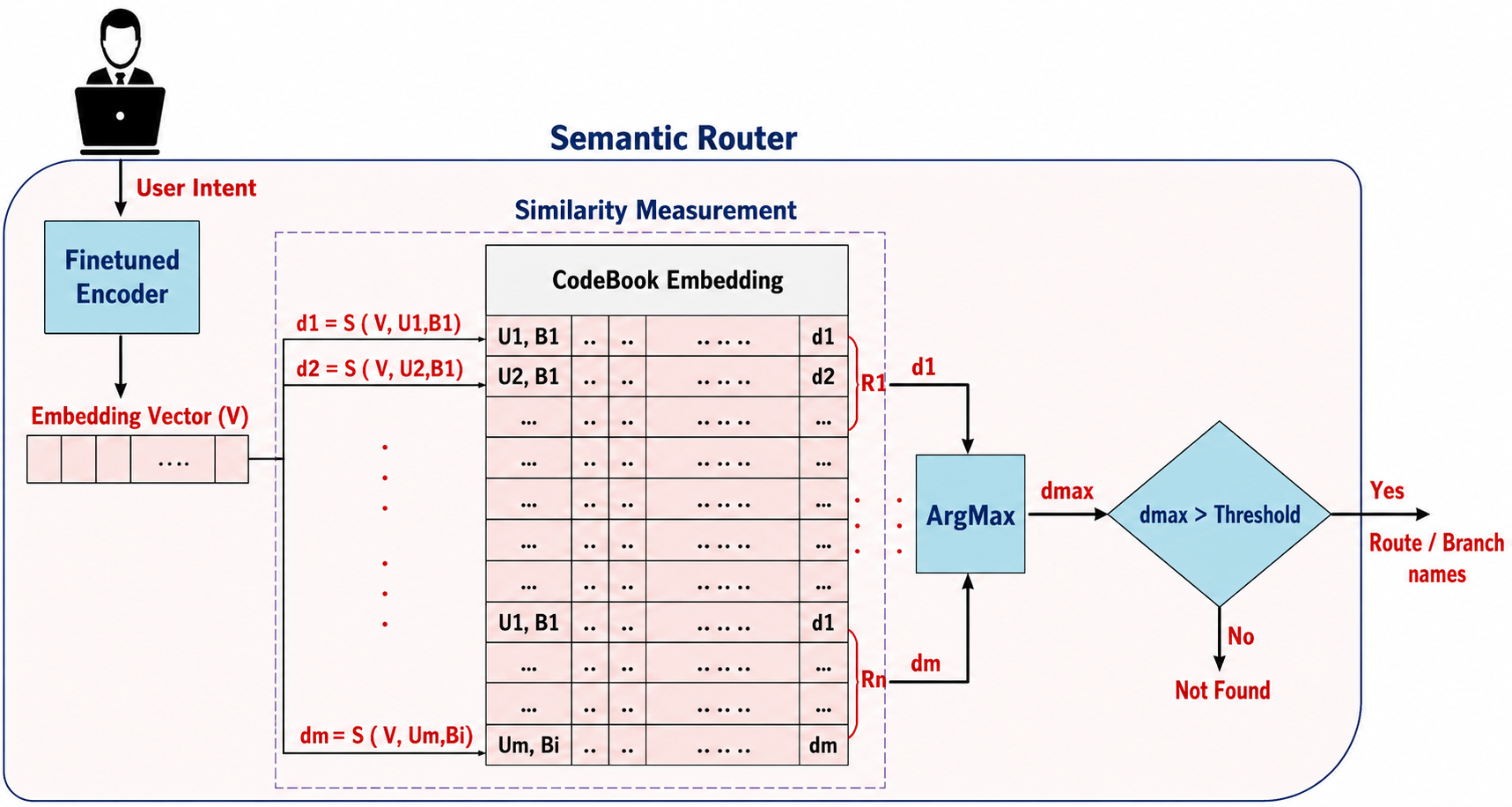}
\caption{Semantic router for encoder-based intent classification: the user intent is encoded and
matched against precomputed codebook embeddings using cosine similarity, and the
route/branch with the highest similarity score is returned.}
\label{Fig:semantic}
\end{figure*}
%%%%%%%%%%%%%%%%%%%%%%%%%%%

\begin{equation}
P_{ij} = \frac{\exp S(V_{A_i}, V_{P_j})}{\sum_{k=1}^{B} \exp S(V_{A_i}, V_{P_k})},
\label{eq:soft}
\end{equation}

\begin{equation}
L = -\frac{1}{B}\sum_{i=1}^{B} \log P_{ii}.
\label{eq:mnrl}
\end{equation}
Finally, the gradients of $L$ are backpropagated to update the shared encoder parameters $\Phi$.
By repeating this process over all mini-batches and epochs, the pretrained encoders are adapted to produce IBN-specific embedding representations. After fine-tuning, the fine-tuned encoders are then utilized in the semantic
router to classify the input intent to the suitable route and
branch.

\subsection{Semantic Router-based Intent Classification}
Following the construction of the telecom-specific codebook, datasets, and the fine-tuning of the selected sentence-transformer encoders, these components are integrated into the semantic router to perform intent classification. The semantic router is a lightweight classification mechanism that maps an input query to a predefined category by measuring its semantic similarity to a set of reference
utterances stored in a codebook~\cite{manias2024semantic}. Formally, given an input intent $q$ and a codebook $\mathcal{C} = \{(u_m, \ell_m)\}_{m=1}^{M}$, where $u_m$
is a reference utterance and $\ell_m$ is its associated label, the semantic router
assigns to $q$ the label of the most similar codebook entry. See
equation~(\ref{eq:SR}), where $f(\cdot)$ is an encoder function that maps text to a dense embedding vector, and
$S(\cdot,\cdot)$ is a similarity measure, typically cosine similarity. 
\begin{equation}
\hat{\ell}(q) = \ell_{\hat{m}}, \quad
\hat{m} = \arg\max_{m} \, S\!\left(f(q),\, f(u_m)\right),
\label{eq:SR}
\end{equation}

To classify the user intent, as illustrated in Fig.~\ref{Fig:semantic}, the fine-tuned encoder is first
used to generate the embedding vectors (\textit{Vs}) for all utterances
contained in the codebook, and these \textit{Vs} are then stored in the
codebook for subsequent similarity comparison. Each stored
\textit{V} is associated with a specific branch \textit{Bi} under
a specific route \textit{Rn}, so that every codebook utterance is linked to its corresponding route/branch label. After that, a new user intent is provided as an input and it is encoded into a vector \textit{V}, which is then compared against all precomputed codebook embeddings using a similarity function. For a given utterance \textit{Um} that belongs to branch \textit{Bi}, the similarity score is computed as illustrated in equation (\ref{eq:similarity}), where $d_m$ is the similarity score between \textit{V}  and the $m$-th codebook utterances.
\begin{equation}
d_m = S(V, V_{u_m}), \quad \forall m,
\label{eq:similarity}
\end{equation}
 The argmax of $\{d_m\}$ is then used to identify the most similar codebook entry, and
the corresponding route/branch label is selected as the classification output, subject
to a confidence threshold $\tau$ as shown in equation (\ref{eq:thre}).
\begin{equation}
\text{Output} =
\begin{cases}
\text{Route/Branch Name}, & \text{if } d_{\max} > \tau, \\
\text{Not Found (NF)}, & \text{otherwise.}
\end{cases}
\label{eq:thre}
\end{equation}
The threshold $\tau$ filters low-confidence predictions and prevents the system from
returning spurious classifications when the input intent does not align with any known
management category. Hence, if the maximum similarity is greater than  $\tau$, the
route and branch labels corresponding to the best-matching codebook utterance are returned as the output. Otherwise, the
input intent is classified as Not Found.

After classifying the user intent, we evaluate the practical deployment efficiency of the proposed phase 1 by comparing the inference time across all encoder types. The total inference time is decomposed as shown in equation (\ref{eq:inference_time}), where $T_{\text{encoding}}(q)$ is the time to generate the embedding of input intent $q$ and $T_{\text{similarity}}(q)$ is the time to match the query embedding against the codebook via cosine similarity. Then, the classified intent goes to downstream steps of policy
translation, validation, and deployment.
\begin{equation}
T_{\text{total}}(q) = T_{\text{encoding}}(q) + T_{\text{similarity}}(q)
\label{eq:inference_time}
\end{equation}
%%%%%%%%%%%%%%%%%%%%%%%%%%%%%%%%%%%%%%%%%%
\section{Phase 2: Intent Translation}
\label{sec:translation}

\begin{figure*}[t]
\includegraphics[width=\linewidth]{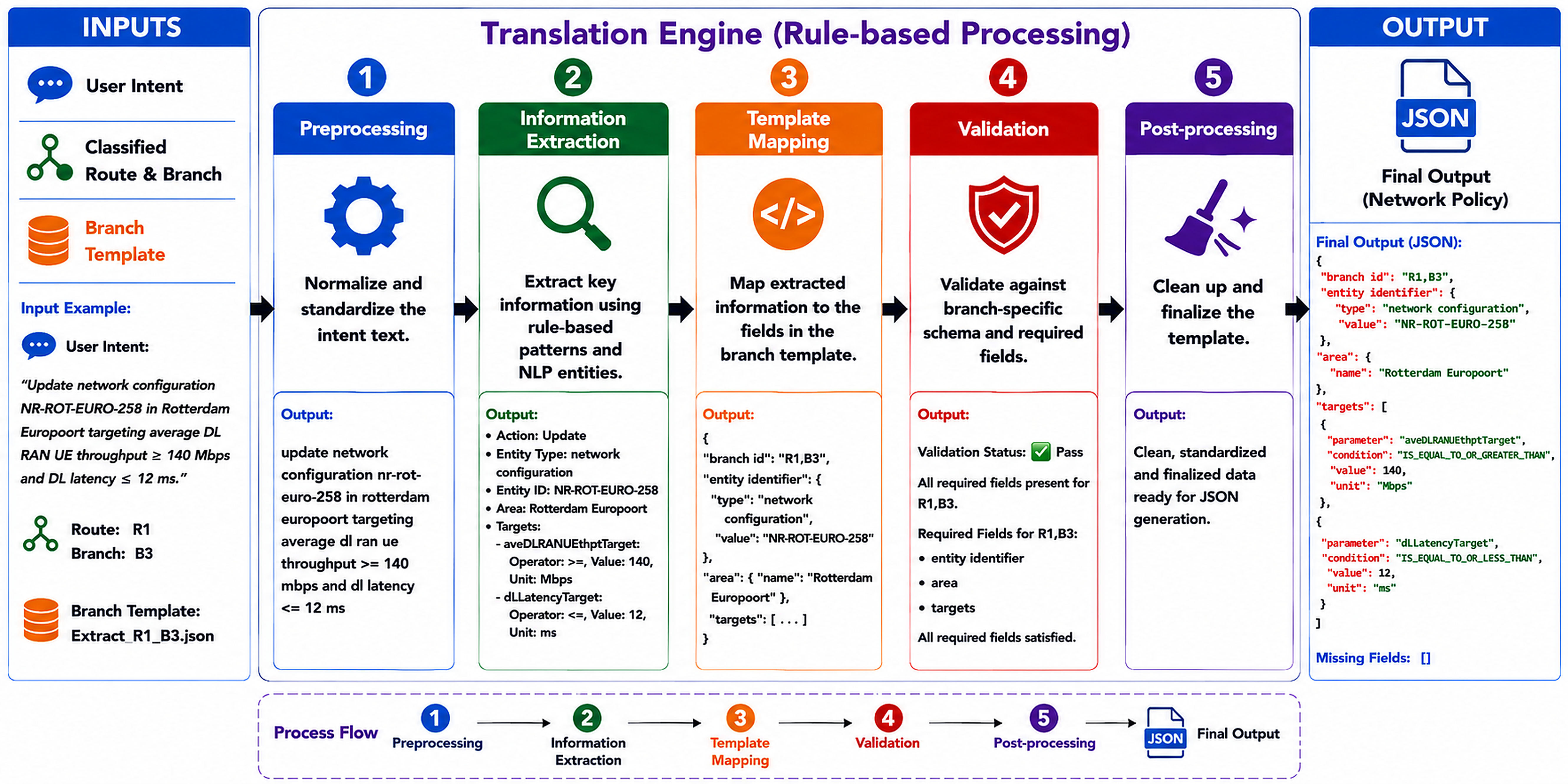}
\caption{Rule-based translator from user intent-to-network policy generation.}
\label{Fig:intent_translation_flow}
\end{figure*}
Following Phase~1, the proposed framework translates the classified intent into a structured network policy that can be directly interpreted by the downstream network management system. This section presents the proposed intent translation mechanism, which is organized into two main components. The first component is a branch-aware, template-driven design that defines the target policy structure for each management operation. While the second component is a hybrid extraction pipeline that identifies, validates, and populates the required policy attributes to generate the final network policy file as an output. 
This design is motivated by the need for deterministic and verifiable outputs in operational telecom systems, where hallucinated or incorrectly extracted
parameters can result in invalid or erroneous network configurations
\cite{wei2025inta,asif2025evaluating}.
%%%%%%%%%%%%%%%%%%%%%%%%%%%
\subsection{Template Design}
%The template-driven design establishes the structural foundation of the proposed translation phase by defining the schema associated with each route/branch pair. Each of the 22 \textit{branches} in the taxonomy is associated with a dedicated JSON template that specifies the complete set of mandatory and optional attributes for that management operation. These templates are derived from the intent-driven management use cases in the 3GPP TS~28.312 and encode domain knowledge about which attributes are structurally required for each branch. For example, an entity identifier and performance targets are mandatory for the \textit{configuration creation intent} (\textit{R1,~B1}), while temporal context, actions, and targets are required for a \textit{self-healing configuration intent} (\textit{R4,~B2}). Once the semantic router assigns an intent to a route/branch pair, the corresponding template is activated and determines the exact fields that must be populated. This branch-gating mechanism has two important advantages. First, it reduces the effective extraction problem to only the parameters that are relevant to the identified management context, eliminating the possibility of generating attributes that do not correspond to any valid field in the branch schema. Second, it enables explicit missing-field detection. So, if a mandatory attribute cannot be extracted from the intent text, the system reports the absent field rather than proceeding with an incomplete output, ensuring that all downstream operations receive complete and  structured representations.
In the translation phase, we create a branch-aware template in which each of the 22 branches in the proposed taxonomy is associated with a dedicated
JSON template. Once the semantic router assigns an input intent to a specific
route/branch pair, the corresponding template is activated and defines the
attributes to be extracted from the intent. Each template specifies the
mandatory and optional attributes required for its associated management operation. For example, the template for the \textit{configuration creation
intent} (\textit{R1,B1}) defines the entity identifier and performance targets
as mandatory attributes, whereas the \textit{self-healing configuration intent}
(\textit{R4,B2}) requires temporal context, actions, and targets. The extracted
information is then mapped to the corresponding fields of the activated template to generate the structured representation of the input intent.

The templates are derived from the intent-driven management use cases defined
in 3GPP TS~28.312 (V18.8.0) and identify the domain knowledge associated with the requirements of each management operation. By activating the template only after the route/branch has been identified, the extraction process is
constrained to the attributes relevant to the selected management context.
This reduces the effective extraction space and prevents the generation of attributes that are not defined for the corresponding branch. In addition,
the distinction between mandatory and optional attributes enables explicit missing-field detection. If a mandatory attribute cannot be extracted from
the input intent, the corresponding field is reported as missing rather than
allowing the system to proceed with an incomplete representation. This
ensures that the resulting structure can be checked against the branch-specific requirements before being passed to the downstream validation and deployment stages.
%%%%%%%%%%%%%%%%%%%%%%%%%%%%%%%%%%
\subsection{Intent Translation Workflow}
The proposed translation workflow consists of three main stages: input processing, translation engine, and Structured outputs. As illustrated in Fig.~\ref{Fig:intent_translation_flow}, the
translator receives three inputs through the input processing stage: the user intent, the classified route/branch
from the semantic router, and the branch-specific template repository. The
translation engine stage then normalize the user intent to standardize its textual
representation. This step resolves Unicode variants, such as Unicode minus
signs and non-breaking spaces, expands unit abbreviations, and normalizes
comparison operators. The normalized intent is subsequently processed by a
hybrid information extraction module that combines rule-based pattern
matching with NLP-based entity recognition. The rule-based component employs
regular expressions to extract structured information that follows predictable
syntactic patterns, including numerical values and \ac{kpi} thresholds, temporal expressions such as date ranges, time windows, and relative durations,
comparison operators mapped to their 3GPP representations, and normalized units. Complementing the
rule-based component, the NLP component utilizes spaCy's
\textit{Matcher} and \textit{EntityRuler} \cite{montani2023explosion, spaCyDocs} to identify domain-specific entities and semantic patterns, including \ac{kpi} names and aliases,
entity identifiers, geographic scope
expressions, and branch-specific condition
clauses.

The extracted attributes are then mapped to a canonical representation to ensure consistency across different natural-language expressions. A
canonical mapping table translates natural-language \ac{kpi} names and
aliases into their corresponding standardized 3GPP attribute names. For
example, \textit{``DL throughput''} is mapped to
\textit{aveDLRANUEthptTarget}, while \textit{``weak-coverage ratio''} is
mapped to \textit{weakRSRPRatioTarget}. The resulting attribute--value pairs
are subsequently forwarded to the template mapping stage, where the branch-specific template corresponding to the classified route and branch is
selected and populated with the extracted information.

The populated template is then validated against its associated branch schema to verify that all mandatory attributes are present and correctly
positioned. If a required attribute cannot be extracted from the input intent, the corresponding missing field is explicitly identified and reported rather than allowing the translator to generate an incomplete
structured representation. Following validation, a post-processing stage
performs template cleanup and formatting to ensure compliance with the target
output schema. Finally, the structured output generation stage produces a
JSON document containing the selected route, branch, and extracted attribute--value pairs as a network policy specification for consumption by
the downstream network infrastructure.

For clarity, Fig.~\ref{Fig:intent_translation_flow} illustrates an example of the full translation process, including the input elements, the intermediate outputs
generated by the translation engine, and the final structured network policy.

In addition, we evaluate the translation performance using a field-level accuracy metric. For each intent, the predicted structured output is compared against its corresponding ground-truth record on an attribute-by-attribute basis. Let $g_{i,j}$ and $p_{i,j}$ denote the ground-truth and predicted values, respectively, for attribute position $j$ in intent $i$. The translation accuracy for a single intent is computed as shown in equation (\ref{eq:Tacc}), where $N_i$ is the total number of evaluated attributes for intent $i$, and $\mathbb{I}(\cdot)$ is the indicator function that equals 1 when the predicted value exactly matches the ground-truth value and 0 otherwise.

\begin{equation}
\text{Acc}_i
=
\frac{\sum_{j=1}^{N_i}
\mathbb{I}\!\left(g_{i,j}=p_{i,j}\right)}
{N_i}
\label{eq:Tacc}
\end{equation}
Therefore, missing values, incorrect values, and misplaced values are all counted as errors. So, the overall translation accuracy is computed as the percentage of correctly predicted attribute values across all evaluated intents.
% ============================================================
\section{Numerical Results}
\label{sec:experiments}
This section presents the experimental setup of the datasets and encoders, as well as the results of the proposed framework.
\subsection{Experimental Setup}
%\subsubsection{Dataset and Codebook Configuration}

Following the procedure described in Section~\ref{sec:dataset}, we utilize the OpenAI GPT-5.2 model using prompt engineering to generate ten base intents for each branch. We use this model since \acp{llm} have demonstrated strong instruction-following capabilities and they have the ability to generate semantically coherent and linguistically diverse text \cite{ouyang2022training}. Then, from the  total of 1,100 intents across the 22 branches, 900 intents are assigned to the route--branch codebook, while the remaining 200 intents form the \ac{id} test set. The \ac{id} test set contains intents drawn from the same route and branch, and linguistic distributions represented in the codebook. To assess the models' generalization capability, an additional set of 200 \ac{ood} intents is constructed using previously unseen base and linguistic diversity intents that are not represented in the codebook. %So, the \ac{id} and \ac{ood} datasets are used to evaluate the encoder-based classification performance.

%\subsubsection{Fine-tuning and Encoder Configurations}
For intent classification experiments, five encoders are evaluated using the same semantic routing framework: the benchmark, and the proposed fine-tuned versions of \texttt{all-mpnet-base-v2} and \texttt{all-MiniLM-L6-v2}, as well as the OpenAI embedding model \texttt{text-embedding-ada-002}. These models are denoted as Mpnet (Base), Mpnet (Tuned), MiniLM (Base), MiniLM (Tuned), and OpenAI, respectively. Each base encoder is fine-tuned twice, once with \ac{mnrl} on  \texttt{Anchor$-$Positive$-$Pairs.jsonl} dataset and once with Triplet loss on the
\texttt{Anchor$-$Positive$-$Negative$-$Pairs.jsonl} dataset, yielding four
fine-tuned variants in total. Both datasets contain 53,900 training samples, following equation (\ref{eq:pair_count}). Additionally, the fine-tuning is performed in Python~3.12.0 using PyTorch~2.11.0 and \texttt{sentence-transformers} library, on both CPU and GPU, using a dual-thread Intel Xeon CPU at 2.20 GHz and an NVIDIA T4 GPU. For fine-tuning, we set the learning rate to $2\times10^{-5}$, 10~epochs, and a batch size
of~64. 
%At inference time, each incoming intent embedding is matched against the precomputed codebook embeddings using cosine similarity, and the route/branch label corresponding to the maximum similarity score is returned as the classification decision.

For intent translation experiments, as a comparative baseline, we evaluate the proposed rule-based translator against the OpenAI GPT-4.1-mini model under identical experimental conditions on 120 intents, 60~base intents and 60~linguistic variational intents, drawn from the same 22-branch taxonomy. The GPT family is selected as a baseline since it has been widely adopted in recent LLM-based \ac{ibn} studies~\cite{manias2024towards, mekrache2024intent}, thereby providing a relevant and reproducible reference point for comparison. In particular, GPT-4.1-mini is chosen due to its strong natural language understanding and instruction-following capabilities while maintaining a lower computational cost and inference latency compared to larger variants~\cite{Open}. To ensure a fair comparison, GPT-4.1-mini receives the same three inputs provided to
the rule-based translator: the input intent, the predicted route/branch
label, and the activated JSON template. Also, the GPT-4.1-mini is augmented with a structured system prompt
encoding the canonical \ac{kpi} name table, condition operator strings, and unit normalization rules. The model
response is constrained to JSON format through the API \texttt{response\_format}
parameter, and the resulting JSON is aligned to the branch schema using the same \texttt{enforce\_template} post-processing applied internally by the rule-based translator.

\subsection{Results}

At the outset, we aim to identify the most suitable loss function to guide the subsequent experiments. To this end, we select the optimal contrastive loss for the proposed semantic routing mechanism  by fine-tuning both
Mpnet and MiniLM encoders with \ac{mnrl} (L1) and Triplet loss (L2),
and evaluate their performance on the \ac{id} and \ac{ood} test sets.

\begin{figure}[t]
\includegraphics[width=\linewidth]{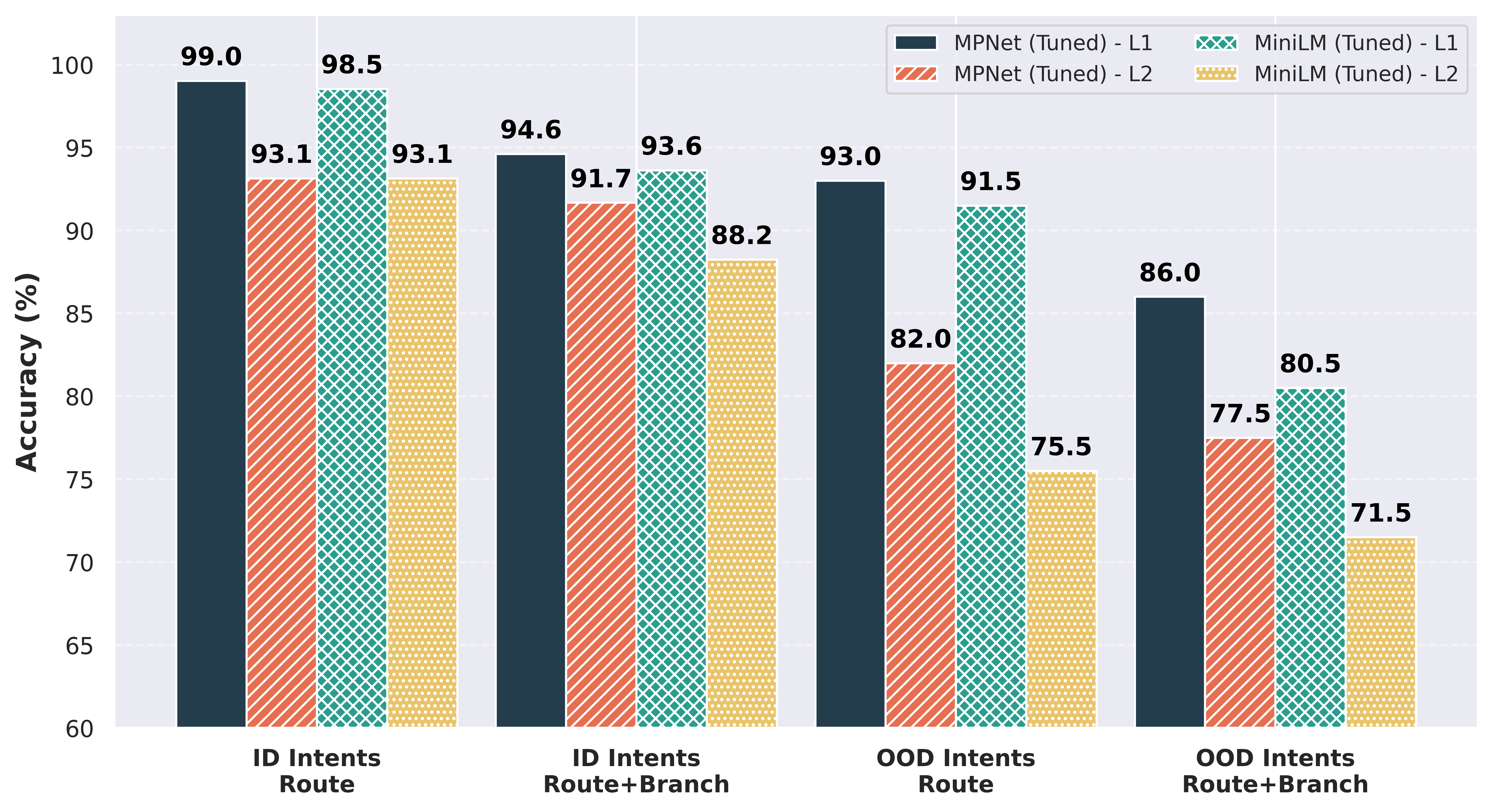}
\caption{Classification accuracy of fine-tuned encoders using MNRL (L1) and Triplet
loss (L2) on 200 in-distribution (ID) and 200 out-of-distribution (OOD) intents.}
\label{fig:loss}
\end{figure}

As reported in  Fig.~\ref{fig:loss},
L1 consistently outperforms L2 across both evaluation settings. For \ac{id} intents,
Mpnet~(Tuned)-L1 achieves 99.02\% route accuracy and 94.61\% route+branch accuracy,
and MiniLM~(Tuned)-L1 achieves 98.53\% and 93.63\%, respectively. While L2 leads
to a noticeable degradation at the branch level, with MiniLM~(Tuned)-L2 dropping to
88.24\%. For \ac{ood} intents, all models experience a performance reduction, as expected from
the increased linguistic diversity of the test set. However, L1 maintains a clear
advantage since the Mpnet~(Tuned)-L1 achieves 93.00\% route accuracy and 86.00\% route+branch
accuracy, while the L2 variants exhibit more pronounced degradations, particularly for MiniLM~(Tuned)-L2. The superior performance of \ac{mnrl} can be attributed to its in-batch negative
strategy, which exposes the encoder to a larger and more diverse set of competing
intents per training step, leading to more discriminative embeddings. In contrast,
Triplet loss relies on explicitly assigned negative pairs and is sensitive to the
quality of triplet selection, which can result in uninformative gradients or unstable
optimization. These observations are consistent with prior findings in contrastive
learning for visual representations~\cite{chen2020simple}, and confirm that in-batch
negatives produce more informative training signals in retrieval-oriented tasks. Based on these results, all subsequent experiments use fine-tuned encoders trained with \ac{mnrl}.
%%%%%%%%%%%%%%%%%%%%%%%%%%%%%%
\begin{figure*}[t]
    \centering
    \begin{subfigure}[t]{0.48\textwidth}
        \centering
        \includegraphics[width=\linewidth]{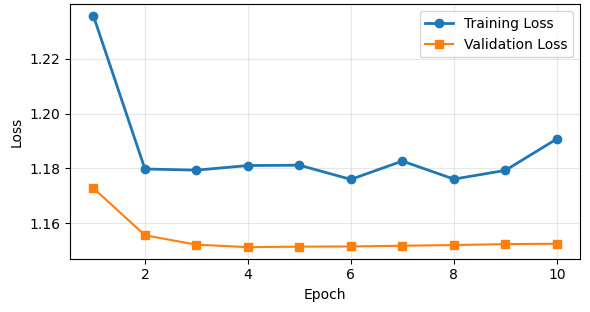}
        \caption{Mpnet (Tuned).}
        \label{fig:mp-val}
    \end{subfigure}
    \hfill
    \begin{subfigure}[t]{0.48\textwidth}
        \centering
        \includegraphics[width=\linewidth]{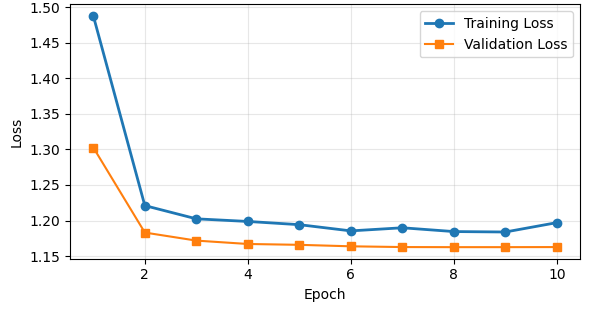}
        \caption{MiniLM (Tuned).}
        \label{fig:mi-val}
    \end{subfigure}
    \caption{Training and validation loss curves across epochs for the two fine-tuned
    encoders.}
    \label{fig:loss-curves}
\end{figure*}

After selecting the contrastive loss function, we proceed to analyze the training behavior and convergence characteristics of the fine-tuned encoders. Fig.~\ref{fig:loss-curves} presents the training and validation loss curves across
10 epochs for both fine-tuned encoders. Both Mpnet and MiniLM exhibit rapid
convergence in the early epochs, with training and validation losses stabilizing
thereafter. A consistent observation across both models is that the validation loss is
lower than the training loss throughout fine-tuning. This behavior is characteristic
of contrastive learning with in-batch negatives because during training, each sample is
contrasted against a large number of competing examples within the batch, creating a
harder optimization problem that yields higher training loss values compared to the
less constrained validation setting~\cite{chen2020simple}.
\begin{table*}[h]
\centering
\caption{Accuracy of the five encoders for 200 in-distribution (ID) and 200 out-of-distribution (OOD) intents.}
\label{tab:seen_unseen_5enc}
\footnotesize
\renewcommand{\arraystretch}{1.1}

\begin{tabular*}{\textwidth}{@{\extracolsep{\fill}} llccccc}
\toprule
\textbf{Dataset} & \textbf{Metric} & \textbf{Mpnet (Base)} & \textbf{MiniLM (Base)} & \textbf{Mpnet (Tuned)} & \textbf{MiniLM (Tuned)} & \textbf{OpenAI} \\
\midrule
\multirow{2}{*}{\textbf{ID Intents}} 
& Route          & 91.67\% & 90.20\% & \textbf{99.02\%} & 98.53\% & 95.59\% \\
& Route + Branch & 85.78\% & 84.31\% & \textbf{94.61\%} & 93.63\% & 87.75\% \\
\midrule
\multirow{2}{*}{\textbf{OOD Intents}} 
& Route          & 42.00\% & 25.00\% & \textbf{93.00\%} & 91.50\% & 60.00\% \\
& Route + Branch & 28.00\% & 12.00\% & \textbf{86.00\%} & 80.50\% & 46.50\% \\
\bottomrule
\end{tabular*}
\end{table*}
Both encoders reach their minimum validation loss at epoch~8, which is selected as the
optimal checkpoint. These results confirm that the proposed fine-tuning strategy
effectively adapts the pretrained encoders to the IBN-specific embedding space within
a modest number of training epochs.

%%%%%%%%%%%%%%%%%%%%%%%%%%%%%%%5
\subsubsection{Intent Classification Results}
\label{sec:accuracy}
After analyzing the behavior of the fine-tuned encoders, we evaluate their classification performance in terms of route level and route+branch-level accuracy for the five encoders on \ac{id} and \ac{ood} test sets.
As shown in table~\ref{tab:seen_unseen_5enc},  for \ac{id} intents, fine-tuning yields clear improvements over the corresponding base
encoders for both model variants. Mpnet~(Tuned) achieves the best overall performance
at 99.02\% route accuracy and 94.61\% route+branch accuracy. MiniLM~(Tuned) achieves
comparable results at 98.53\% and 93.63\% for route and branch classification, respectively. The OpenAI encoder achieves
intermediate performance, outperforming both base encoders but underperforming compared to
fine-tuned models, indicating that domain adaptation through contrastive fine-tuning
provides a measurable benefit even relative to a strong general-purpose encoder. For \ac{ood} intents, a more pronounced performance divergence is observed across the
five encoders. The base encoders exhibit a sharp accuracy drop: MiniLM~(Base) falls to
25.00\% route accuracy and 12.00\% route+branch accuracy. The OpenAI encoder
generalizes more effectively, achieving 60.00\% and 46.50\%, respectively, reflecting
the benefits of large-scale pretraining on diverse text corpora. Nevertheless, the
fine-tuned encoders yield substantially higher accuracies with Mpnet~(Tuned) achieving 93.00\%
route accuracy and 86.00\% route+branch accuracy, demonstrating robust generalization
to \ac{ood} for base intents and linguistic formulations.

\begin{figure*}[t]
    \centering
    \includegraphics[width=.9\linewidth]{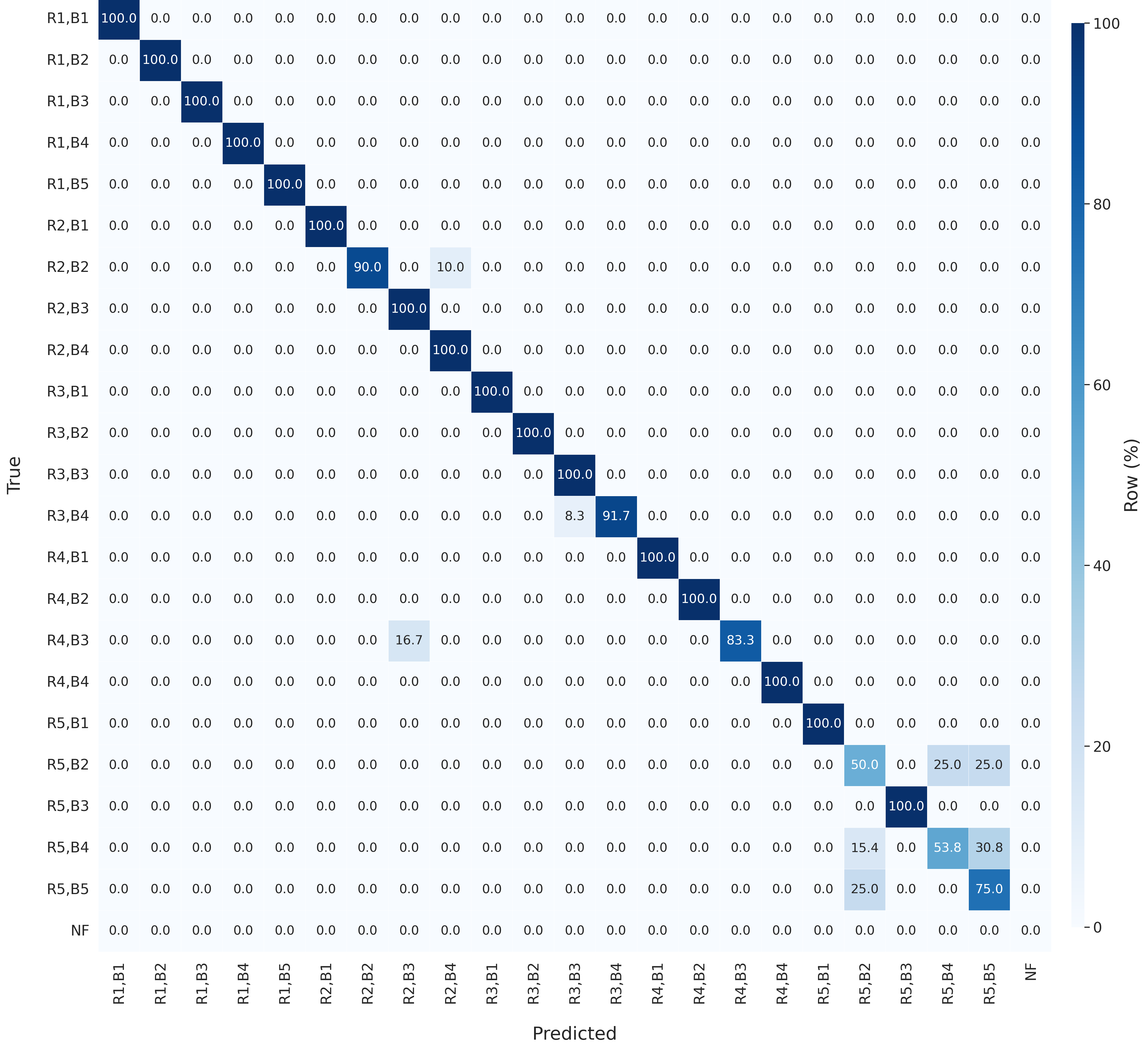}
    \caption{Confusion matrix for the five routes using the Mpnet~(Tuned) encoder on
    the full test set.}
    \label{fig:tuned1}
\end{figure*}

\begin{figure*}[t]
\centering
\begin{subfigure}[t]{0.22\textwidth}
\centering
\includegraphics[width=\linewidth]{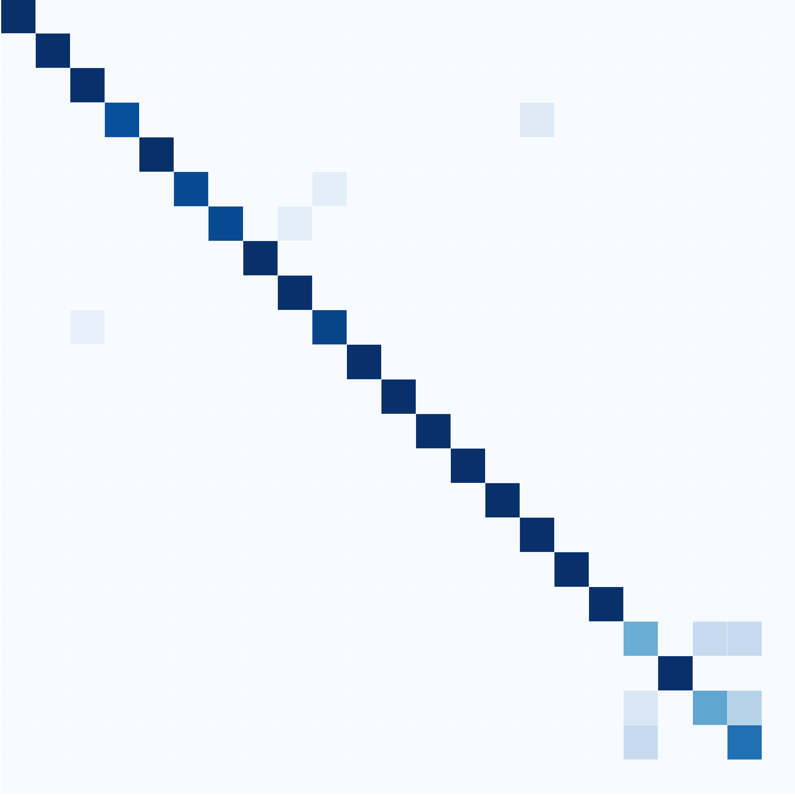}
\caption{MiniLM~(Tuned).}
\label{fig:tuned2}
\end{subfigure}
\hfill
\begin{subfigure}[t]{0.22\textwidth}
\centering
\includegraphics[width=\linewidth]{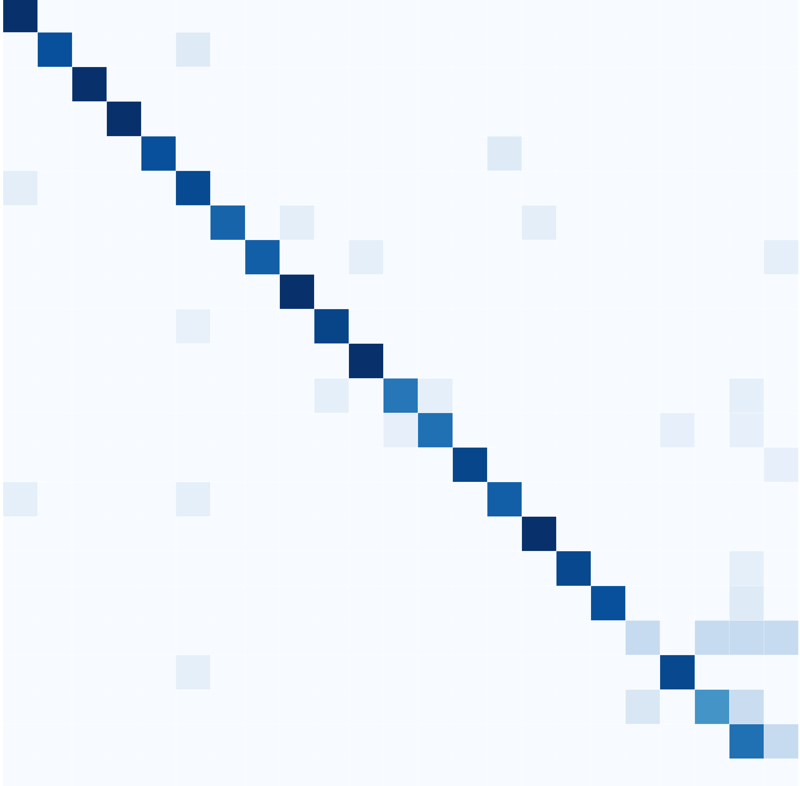}
\caption{Mpnet~(Base).}
\label{fig:base1}
\end{subfigure}
\hfill
\begin{subfigure}[t]{0.22\textwidth}
\centering
\includegraphics[width=\linewidth]{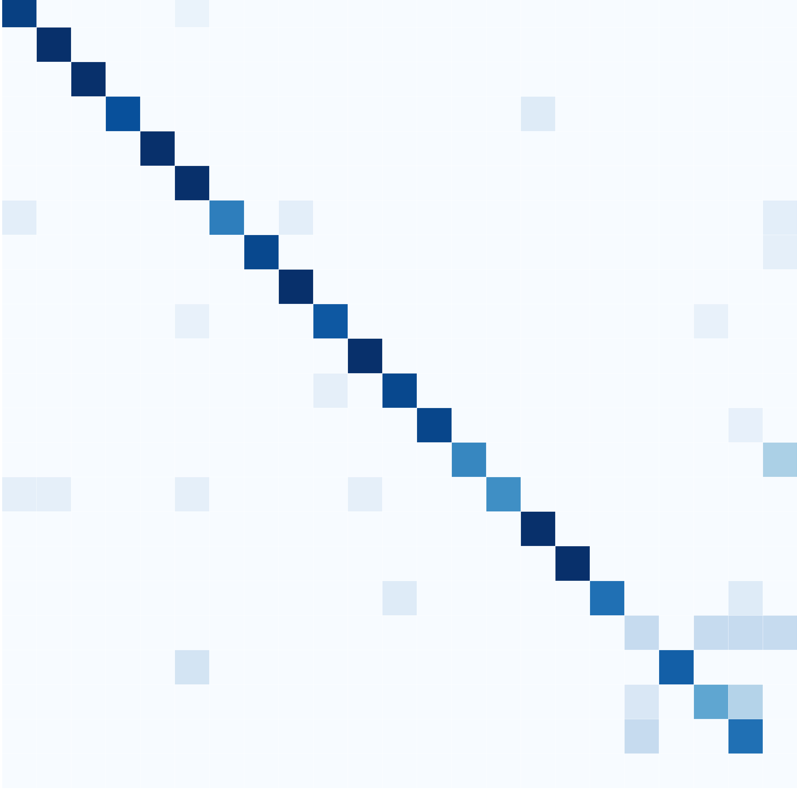}
\caption{MiniLM~(Base).}
\label{fig:base2}
\end{subfigure}
\hfill
\begin{subfigure}[t]{0.26\textwidth}
\centering
\includegraphics[width=\linewidth]{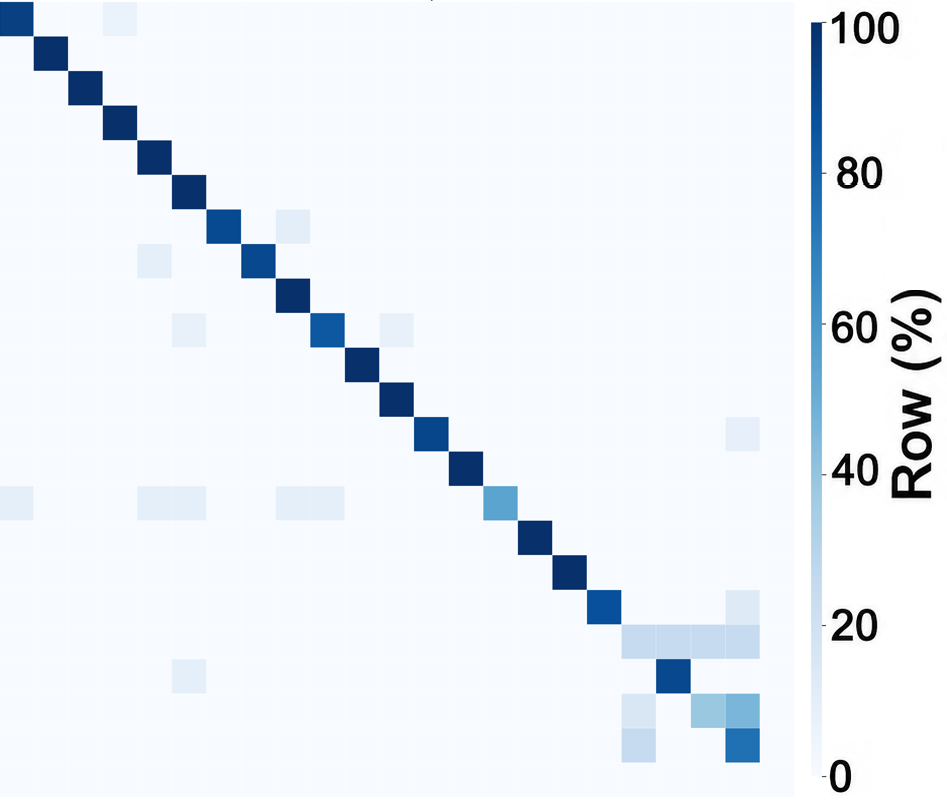}
\caption{OpenAI.}
\label{fig:openAI}
\end{subfigure}
\caption{Confusion matrices for the five routes across the four remaining encoders.}
\label{fig:cm-all}
\end{figure*}

Confusion matrices for all five encoders are presented in Figs.~\ref{fig:tuned1}
and ~\ref{fig:cm-all}, where rows correspond to true labels and columns to predicted
labels. In Fig.~\ref{fig:tuned1}, Mpnet~(Tuned) achieves near-perfect
classification for the majority of the  branches, with several diagonal entries reaching
100\% accuracy (e.g., \textit{R1: B1--B5}, \textit{R2: B1}, \textit{R3: B1--B3}, \textit{R4: B1--B2}, and \textit{R5: B1--B3}). \textit{Routes R1} through \textit{R4} are handled accurately by all fine-tuned encoders while \textit{R5
(Network Analytics \& Operational Reporting)} is the most challenging across all
models, particularly at the branch level. The branches in  \textit{R5} such as \textit{ B2 (metric retrieval), B4
(performance reporting)}, and \textit{B5 (configuration/service listing)}, share significant
semantic overlap. They frequently co-occur with similar action verbs (e.g., ``show'',
``get'', ``retrieve'', ``list'', ``generate'') and overlapping domain-specific terms
(e.g., \acp{kpi}, performance, network, data, configurations). The primary distinction
between these branches often lies in the expected output format, which is a subtle
semantic cue that can be difficult to capture solely through cosine-similarity-based
matching. The base encoders additionally produce Not~Found (NF) predictions, indicating that the
query embedding did not sufficiently match any codebook entry above the confidence
$\tau$. This behavior is visible in Fig.~\ref{fig:base2}, where MiniLM~(Base)
assigns 33.3\% of \textit{R4,B1} predictions to NF and 25\% for \textit{R5,B2}. A similar pattern is
observed for Mpnet~(Base) in Fig.~\ref{fig:base1}. In contrast, the fine-tuned
encoders substantially reduce NF occurrences, reflecting improved alignment between
the embedding space and the codebook representation. Overall, the fine-tuned encoders provide the most accurate and stable branch-level
predictions. Among all evaluated models, Mpnet~(Tuned) achieves the highest
performance across all routes and branches, followed by MiniLM~(Tuned), then the
OpenAI encoder, while the base encoders consistently exhibit lower and less reliable
accuracy.

After evaluating the classification performance of the proposed encoders, we  analyze their latency to assess their suitability for real-time deployment scenarios. This step is essential to understand the trade-off between inference latency and classification accuracy. 
\begin{table}[t]
\centering
\caption{Average inference time per intent for different encoders under CPU and GPU execution in ms.}
\label{tab:inference_time}
\scriptsize
\renewcommand{\arraystretch}{1.0}
\setlength{\tabcolsep}{3pt}
\resizebox{\columnwidth}{!}{%
\begin{tabular}{llccc}
\toprule
\textbf{Execution in (ms)} & \textbf{Encoder} & \textbf{$T_{\text{query\_encoding}}$} & \textbf{$T_{\text{similarity\_measurement}}$} & \textbf{$T_{\text{total}}$} \\
\midrule
\multirow{3}{*}{\textbf{CPU}}
& Mpnet  & 177.705 & 0.529 & 178.234 \\
& MiniLM & 27.617  & 0.194 & \textbf{27.811} \\
& OpenAI & 209.270 & 2.871 & 212.141 \\
\specialrule{.12em}{.08em}{.08em}
\multirow{3}{*}{\textbf{GPU}}
& Mpnet  & 11.529  & 0.159 & 11.688 \\
& MiniLM & 5.093   & 0.076 & \textbf{5.168} \\
& OpenAI & 104.635 & 1.436 & 106.071 \\
\bottomrule
\end{tabular}%
}
\end{table}
Table~\ref{tab:inference_time} reports the average per-intent inference time for CPU
and GPU execution.  For the OpenAI encoder, $T_{\text{query\_encoding}}$
includes network latency and remote server processing; for locally deployed encoders,
it corresponds to the forward-pass computation time on the target device. For the four other encoders, the base and fine-tuned versions of each encoder share
identical architectures, their inference times are identical; only Mpnet and MiniLM
are therefore reported separately. MiniLM achieves the lowest inference time, 27.81\,ms
on CPU and 5.17\,ms on GPU. Mpnet requires approximately five times longer inference,
while the OpenAI encoder exhibits the highest latency due to network overhead and
remote processing. These results illustrate a clear accuracy-efficiency trade-off. Mpnet~(Tuned) achieves
the highest classification accuracy, while MiniLM~(Tuned) is substantially faster. The
choice of the encoder should therefore reflect the target deployment context. For example, in
latency-critical applications, such as \ac{urllc} management and industrial automation with closed-loop control~\cite{ji2018ultra}, MiniLM~(Tuned) is the preferred option. While in scenarios where semantic accuracy is paramount and processing time is less
constrained, such as network planning, policy validation, or offline optimization,
Mpnet~(Tuned) provides a superior accuracy profile.

%After evaluating the inference efficiency of the encoder-based classification module, we now assess the performance of the intent translation component in terms of both accuracy and latency.
\subsubsection{Intent Translation Results}
\label{sec:translation_results}

\begin{figure}[t]
\includegraphics[width=\linewidth]{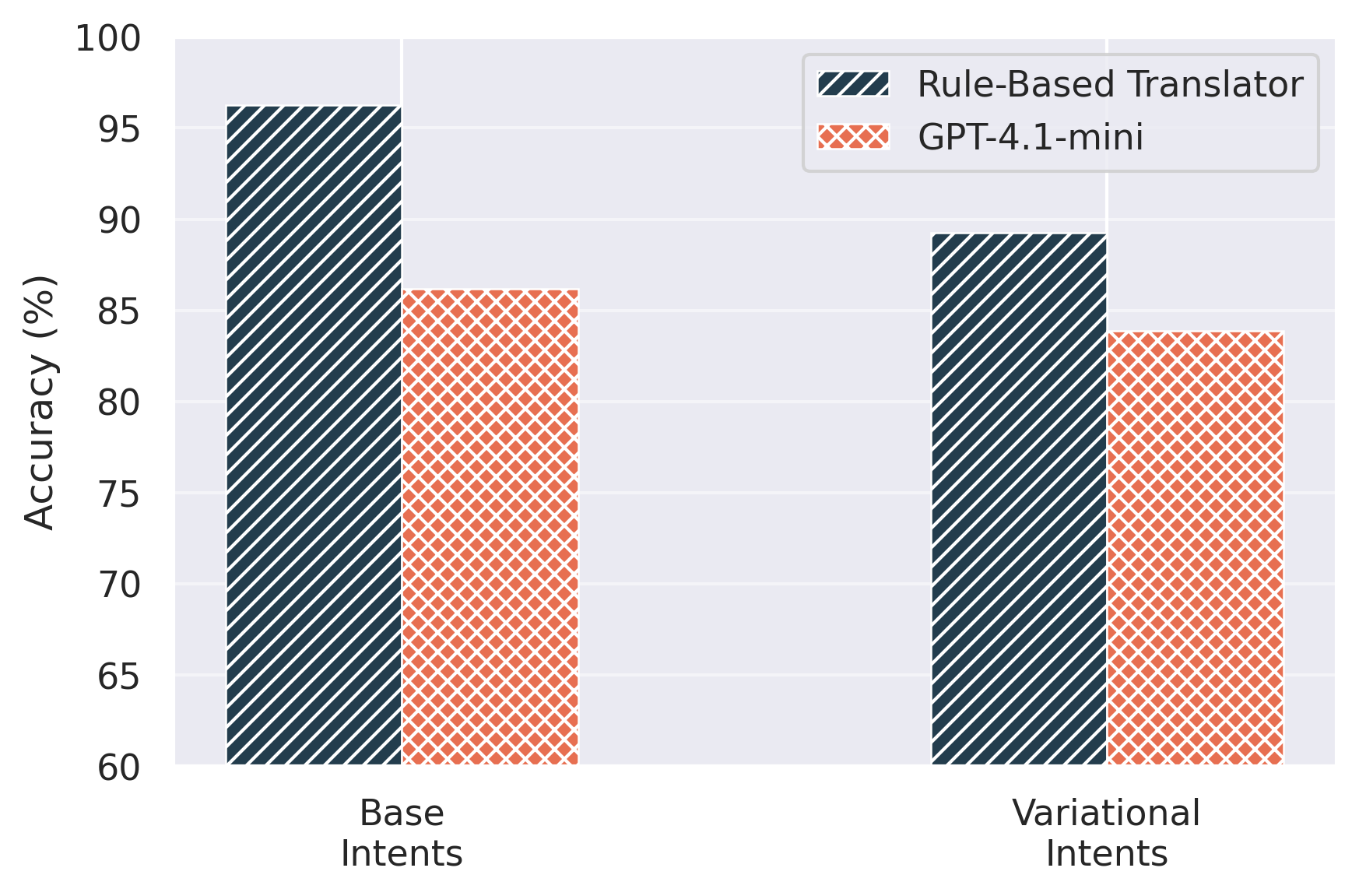}
\caption{Intent translation accuracy of the rule-based translator and GPT-4.1-mini on
60~base intents and 60~variational intents.}
\label{fig:translation_accuracy}
\end{figure}
As illustrated in Fig.~\ref{fig:translation_accuracy}, the rule-based translator achieves
96.24\% overall accuracy on the 60~base intents, compared to 85.37\% for GPT-4.1-mini, a margin of 10.87 percentage points. The rule-based translator’s advantage on base intents is driven by its deterministic structure. It enforces canonical \ac{kpi} and entity naming through a fixed alias table, ensuring exact 3GPP-compliant tokens, whereas GPT-4.1-mini produces paraphrased or semantically equivalent but non-matching field names. It also consistently populates required \textit{null} placeholders in optional schema fields, which are frequently omitted by the LLM. On the 60~variational intents, the translator
achieves 89.26\% versus 82.28\% for GPT-4.1-mini, a reduced but still substantial margin of 6.98 percentage points. 
Hence, the performance degradation is more pronounced for the rule-based system compared to GPT-4.1-mini, indicating higher sensitivity to lexical perturbations in entity identifier, temporal context, and \ac{kpi}-related expressions. In contrast, GPT-4.1-mini exhibits improved robustness due to its semantic generalization capabilities; however, it still presents structured-output inconsistencies. Specifically, the model produces hallucinated values that are not grounded in the input intent, shows inconsistency in canonical naming. Also, it does not consistently represent missing information in the expected output structure. Moreover, the generated outputs are not fully deterministic across multiple runs, resulting in observable variability.

\begin{figure}[t]
    \centering

    \begin{subfigure}[t]{\columnwidth}
        \centering
        \includegraphics[width=0.8\columnwidth]{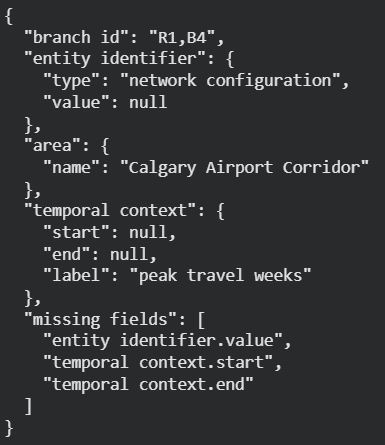}
        \caption{Rule-based Translator output}
        \label{fig:translator-output}
    \end{subfigure}

    \vspace{0.3cm}

    \begin{subfigure}[t]{\columnwidth}
        \centering
        \includegraphics[width=0.8\columnwidth]{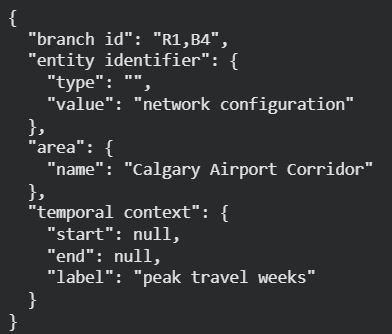}
        \caption{GPT-4.1-mini output}
        \label{fig:gpt-mini-output}
    \end{subfigure}

    \caption{Comparison between translator and GPT-mini outputs.}
    \label{fig:comparison-outputs}
\end{figure}

For more clarification, Fig.~\ref{fig:comparison-outputs} illustrates a comparison between the rule-based translator and GPT-4.1-mini for the following incomplete intent: \textit{``Pause network configuration in Calgary Airport Corridor during peak travel weeks"}. As shown in Fig.~\ref{fig:translator-output}, the rule-based translator extracts all available attributes and explicitly reports the missing ones and produces a fully JSON structured output. In contrast, Fig.~\ref{fig:gpt-mini-output} shows that GPT-4.1-mini incorrectly maps the value of \textit{entity\_identifier.type} to \textit{entity\_identifier.value}. Additionally, although the model identifies missing temporal boundaries (i.e., start and end times), it does not explicitly include them in the designated ``missing fields" section, despite being explicitly specified in the prompt. Furthermore, GPT-4.1-mini exhibits instability across repeated executions for the same input. In some cases, it assigns \textit{null} values despite the presence of extractable information, misassigns attributes (e.g., mapping \textit{area.name} to \textit{entity\_identifier.value}), or alters semantic interpretation of fields, such as replacing ``network configuration" with ``stop configuration" in the \textit{entity\_identifier.value} field.

Beyond accuracy, translation latency is a critical consideration for operational
network management, where closed-loop control cycles impose strict timing budgets. The average translation time per intent using 
rule-based translator is 4.81\,ms, executing
entirely on local hardware without any network overhead. In contrast, GPT-4.1-mini
requires an average of 4{,}163.58\,ms per intent, a factor of approximately
$866{\times}$ longer, attributable primarily to network round-trip latency and remote
server processing time. This latency gap has direct practical implications for deployment. In time-sensitive
scenarios such as \ac{urllc} management~\cite{ji2018ultra}, self-healing loop
activation, or real-time interference mitigation, the sub-10\,ms response time of the
rule-based translator is compatible with closed-loop control requirements, whereas the
multi-second latency of GPT-4.1-mini would violate the timing constraints of such
applications. Furthermore, when considered within the context of the full end-to-end
framework, combining MiniLM~(Tuned) inference on GPU (5.17\,ms) with the rule-based
translator (4.81\,ms) yields a total per-intent pipeline latency of approximately
9.98\,ms, confirming that the proposed system as a whole is suitable for real-time
deployment. Replacing the rule-based translator with GPT-4.1-mini would increase the
total per-intent pipeline latency to approximately 4.17\,s even with the fastest local
encoder, making LLM-based translation impractical for latency-sensitive management
functions.
\section{Conclusion}
\label{sec:conclusion}

This paper presented a semantic routing framework for multi-domain telecom \ac{ibn} intent classification and translation. The proposed framework addresses the limitations of server-based LLM solutions by leveraging locally deployed, fine-tuned sentence-transformer encoders and a 3GPP TS~28.312-aligned route/branch taxonomy for intent classification. A telecom-specific codebook enriched with linguistic diversity methods is introduced to capture variability in natural-language expressions across multiple network domains, including the \ac{ran}, \ac{5g} core, etc. The proposed contrastive fine-tuning strategy enables effective adaptation of the embedding space to the \ac{ibn} domain, improving discrimination between semantically similar branches. In addition, the proposed branch-aware, template-driven translation mechanism ensures deterministic and schema-compliant policy generation while eliminating hallucinated outputs. The numerical results demonstrate that the fine-tuned encoders, especially the Mpnet encoder, significantly outperform both the base encoders and a commercial embedding baseline in both \ac{id} and \ac{ood} testing sets. Moreover, the MiniLM encoder provides a favorable trade-off between accuracy and latency, making it suitable for time-sensitive deployments. For intent translation, the proposed rule-based approach achieves higher accuracy and significantly lower inference latency compared to the commercial-based generation, highlighting its suitability for real-time and resource-constrained environments. Overall, the proposed framework provides a practical, scalable, and fully deployable solution for intent classification and translation in the \ac{ibn} lifecycle for next-generation networks.

%Future work will focus on extending the semantic routing framework to support automated and dynamic codebook expansion for emerging \ac{6g} use cases. In addition, multi-label and hierarchical intent routing will be investigated to better handle cross-domain and cross-layer intents, along with lightweight mechanisms for disambiguating closely related branches.

% ============================================================
\bibliographystyle{IEEEtran}
\bibliography{references}

\end{document}